\documentclass[a4paper,fleqn,usenatbib,useAMS]{mnras}

\usepackage{graphicx}	
\usepackage{amsmath}	
\usepackage{amssymb}	
\usepackage{multicol}        
\usepackage{bm}		
\usepackage{pdflscape}	
\usepackage{mathptmx}

\usepackage[T1]{fontenc}
\usepackage{ae,aecompl}

\title[Horndeski gravity in pulsars]{Horndeski gravity without screening in binary pulsars}
	
\author[P. I. Dyadina et al.]{Polina I. Dyadina,$^{1,2}$\thanks{E-mail: \href{mailto:guldur.anwo@gmail.com}{guldur.anwo@gmail.com}},
	Nikita A. Avdeev,$^{3}$\thanks{E-mail: \href{mailto:guldur.anwo@gmail.com}{naavdeev1995@mail.ru}}
	and Stanislav O. Alexeyev$^{2,4}$ , \thanks{E-mail: \href{mailto:guldur.anwo@gmail.com}{alexeyev@sai.msu.ru}}
\\
$^{1}$Department of Astrophysics and Stellar Astronomy, Faculty of Physics, Lomonosov Moscow State University, \\
Leninskie Gory, 1/2, Moscow 119991, Russia\\
$^{2}$Sternberg Astronomical Institute, Lomonosov Moscow State University, Universitetsky Prospekt, 13, Moscow 119991, Russia\\
$^{3}$Department of Celestial Mechanics, Astrometry and Gravimetry, Faculty of Physics, Lomonosov Moscow State University,\\
 Leninskie Gory, 1/2, Moscow 119991, Russia\\
$^{4}$Department of Quantum Theory and High Energy Physics, Faculty of Physics, Lomonosov Moscow State University,\\
 Leninskie Gory, 1/2, Moscow 119991, Russia}

\date{Accepted XXX. Received YYY; in original form ZZZ}

\pubyear{2018}

\begin{document}
\label{firstpage}
\pagerange{\pageref{firstpage}--\pageref{lastpage}}
\maketitle

\begin{abstract}
	We test the subclasses of Horndeski gravity without Vainshtein mechanism  in the strong field regime of binary pulsars. We find the rate of energy losses via the gravitational radiation predicted by such theories and compare our results with observational data from quasi-circular binaries PSR J1738+0333, PSR J0737-3039, PSR J1012+5307. In addition, we consider few specific cases: the hybrid metric-Palatini f(R)-gravity and massive Brans-Dicke theory. 
\end{abstract}

\begin{keywords}
	 gravitation -- pulsars: general -- gravitational waves -- methods: analytical
\end{keywords}

\section{Introduction}\label{sec:Introduction}

The General Relativity (GR) is the universally recognized theory of gravity. It successfully describes a wide range of scales and gravitational regimes (weak field limit in Solar System and strong field regime of binary black holes). Together with Standard model, they  represent two pillars of modern physics. 

Unfortunately, some phenomena cannot be explained completely in the frameworks of these two approaches. The accelerated expansion of our Universe has been found from the Supernovae Type Ia (SN Ia) observations \citep{acceleration, acceleration1, acceleration2, acceleration3}. So  an extra component called ``Dark Energy'' (DE) has been introduced by \citet{dark energy}, but the nature of this phenomenon is not fully understood. The other problem is dark matter \citep{Zwicky1, Zwicky}. It is the invisible matter, which fills up galaxies and manifests itself only in the gravitational interaction. Also, this phenomenon can be described (apart from ``new physics'') by changing the gravitational theory at galaxy scales \citep{dark matter3, dark matter1, dark matter2, dark matter}. Furthermore, there is no any complete self-consistent quantum theory of gravity. All these facts lead to an increasing number of modified gravitational theories. One of the most widespread approaches to create the modified gravity is to extend GR with higher order curvature corrections and additional degrees of freedom  \citep{Alexeev:1996vs, Alexeyev:2012zz}. But the simplest way to modify GR remains adding of a scalar field. 

The Horndeski gravity is the most general scalar-tensor theory providing the second-order field equations which evades Ostrogradski instabilities \citep{horndeski}. It represents a covariant generalization of Galileon gravity. Horndeski gravity suggests solutions for some GR's problems. For example, the scalar field can play the role of DE and explain the accelerating expansion of the Universe \citep{dark energy in horndeski}. Therefore during last few years in connection with all these circumstances, the Horndeski gravity attracts a large number of researchers. This theory has recently been studied extensively in the context of cosmology \citep{germani, germani1, germani2} and physics of black holes \citep{Dasha1, Dasha}. Taking into account the generality and importance of Horndeski model, it is natural to ask how this theory pass different experimental gravitational tests and impose restrictions on its parameters. The Horndeski gravity has already been tested in many experiments (cluster lensing \citep{narikawa}, the cosmic microwave background (CMB) data \citep{salvatelli, salvatelli1} and so on). Special attention should be paid to the recent works of \citet{ez} and \citet{Baker} related to the verification of the Horndeski theory using LIGO data for event GW170817 \citep{abbott} and the concomitant gamma-ray burst GRB 170817A \citep{grb}. In these papers authors investigate the speed of gravitational waves in various theories and show that data of the binary neutron star merger GW170817 \citep{abbott} and the concomitant gamma-ray burst GRB 170817A \citep{grb} allow to restrict the parameters of the Horndeski gravity.  

The most general form of Horndeski gravity predicts the existence of a fifth force which is strongly constrained by precision tests at Solar System scales. If a theory involves a scalar field for description of DE, it should contain a mechanism for suppressing of the scalar interaction with visible matter on small scales, that it relates only to cosmological scales. The Vainshtein mechanism represents such a possibility \citep{Vainshtein}. Originally it was used in application to massive gravity \citep{Vainshtein}. Now this mechanism is actively applied to the Horndeski models  due to the presence of non-linear derivative interactions \citep{kimura, koyama}. The Vainshtein mechanism claims that it is not possible to ignore the effect of nonlinearity within the so-called Vainshtein radius $r_V$ from the center of the matter source. Beyond $r_V$, the linearization can be applied. In this paper we consider the subclass of Horndeski theories which do not imply Vainshtein mechanism in the strong field regime of binary pulsars. Such subclass of Horndeski gravity reduces to the standard massive scalar-tensor theories. A similar task was investigated in the work of \citet{hou} where authors restrict their consideration with massless case. 

Investigations of other types of screening  mechanisms, such as the chameleon \citep{khw, khw1}, the symmetron \citep{hint, hint1} and the dilaton \citep{polyakov, polyakov2}, in the context of scalar-tensor theories are widespread. The question about screening effects manifestations in binary pulsars data was considered earlier by \citet{brax, zhang}. In this paper we focus only on scalar-tensor models that do not imply any types of screening mechanisms.

The discovery of the first binary pulsar system  PSR B1913 + 16 by \citet{hulse} has opened a new testing ground for GR and its extensions. It is important to emphasize that in binary pulsars one deals with a gravitational field, which is stronger than in the Solar System. Moreover, due to the high stability of the pulse arrival it is possible to extract the dynamics of the orbital motion with such an accuracy at which the effects of gravitational waves emission could appear. The observable orbital decay of binary pulsars became the first experimental proof of the gravitational radiation existence. Now observational data of the orbital period change have a high accuracy. All these facts make the binary pulsars an indispensable laboratory for studying the behaviour of gravitational radiation in different models of gravity \citep{pshirkov}. In addition, pulsars allow to  understand other physical processes better \citep{Ivanov:2016ifg}. 

The scalar dipole radiation dominates in the expressions for the orbital decay of the binary pulsars predicted by scalar-tensor theories \citep{eardley, will, massivebd, zhang}. This dipole contribution to the gravitational radiation is produced due to violations of the gravitational weak equivalence principle (GWEP) \citep{casola}. This effect becomes more pronounced in mixed binaries (binary systems whose members have different gravitational binding energy). The fact is that the dipole radiation is produced when the system's centre of mass is offset with respect to the centre of inertia. So mixed binaries and eccentric ones seems to be the best target to constrain scalar-tensor theories \citep{massivebd}. In this work, we test subclass of Horndeski gravity (without Vainshtein mechanism as a first step) in mixed binary systems and impose restrictions on the parameters of this model.

The structure of the paper is the following. In section \ref{sec:Horndeski gravity} we discuss the action of the Horndeski gravity and reduce it to the standard massive scalar-tensor action. The section \ref{sec:Field equations in the weak-field limit} contains the field equations in the weak-field limit. In section \ref{sec:Post-newtonian solutions}, we solve the post-Newtonian equations for the tensor and scalar fields. Further, in section \ref{sec:EIH equations of motion} we find the motion equations of binary systems. After that, in section \ref{sec:GRAVITATIONAL RADIATION FROM COMPACT BINARIES} we obtain the stress-energy pseudotensor using the Noether current method, calculate the rate of the energy loss due to the tensor and scalar gravitational radiations and derive their contributions to the orbital period change. In section \ref{sec:Observational constrains on Horndeski gravity from binary pulsars}, we derive the constraints on the parameters of  the standard massive scalar-tensor theories by the current observations; also in this section we consider two specific models: the hybrid metric-Palatini f(R)-gravity and massive Brans-Dicke theory, and impose restrictions on these models. We conclude in section \ref{sec:conclusions} with a summary and discussion.

Throughout this paper the Greek indices $(\mu, \nu,...)$ run over $0, 1, 2, 3$ and the signature is  $(-,+,+,+)$. All calculations are performed in the CGS system.

\section{Massive scalar-tensor gravity}
\label{sec:Horndeski gravity}

\subsection{Action}

We start our consideration from the action of the Horndeski theory, which is presented by \citet{kobayashi},
\begin{equation}\label{act}
S = \frac{c^4}{16\pi}\sum_{i=2}^5\int d^4x \sqrt{-g} L_i + \int d^4x L_m(A^2(\phi)g^{\mu\nu}, \Psi_m^{(j)}) ,
\end{equation}
where $c$ is the speed of light, $g$ is the determinant of the metric, and $L_m$ is the Lagrangian density for the matter fields $\Psi_m^{(j)}$ labeled by $j$. $L_i$ are the  gravitational Lagrangian densities:
\begin{eqnarray}\label{Lagrangians}
L_2&=&G_2(\phi, X),\ \  L_3=-G_3(\phi, X)\Box\phi, \nonumber\\
L_4&=&G_4(\phi, X)R+G_{4X}[(\Box\phi)^2-(\nabla_{\mu}\nabla_{\nu}\phi)^2], \nonumber\\
L_5&=&G_5(\phi, X)G_{\mu\nu}\nabla^{\mu}\nabla^{\nu}\phi-\frac{G_{5X}}{6}\biggl[(\Box\phi)^3+2(\nabla_{\mu}\nabla_{\nu}\phi)^3 \nonumber\\
&&-3(\nabla_{\mu}\nabla_{\nu}\phi)^2\Box\phi\biggr],
\end{eqnarray}
where $G_{\mu\nu}$ is the Einstein tensor, $R$ is Ricci scalar, $\phi$ is the scalar field, $X=-1/2\nabla_{\mu}\phi\nabla^{\mu}\phi$, $\nabla_{\mu}$ is the covariant derivative, $\Box\phi=g^{\mu\nu}\nabla_{\mu}\nabla_{\nu}\phi$, $G_i(\phi, X)$ are arbitrary functions of the scalar field $\phi$ and its kinetic term $X$, $G_{iX}=\frac{\partial G_i}{\partial X}$. The choice of the specific type of arbitrary functions determines the particular gravitational theory. 

In this work we consider the matter Lagrangian density which depends on the gravitational fields according to
\begin{equation}
L_m= L_m(A^2(\phi)g^{\mu\nu}, \Psi_m^{(j)})
\end{equation}
where $A(\phi)$ is an arbitrary function of $\phi$. Using the conformal transformation $g_{\mu\nu}\rightarrow A^2(\phi)g_{\mu\nu}$, we move from the Einstein frame to the Jordan one, where the matter fields do not couple directly to the scalar field but the indirect coupling occurs via the metric \citep{Fujii, Esp, Clifton}.

In our work we investigate only subclass of Horndeski gravity without screening mechanisms. The general action reduces to the considering model with the following set of the gravitational Lagrangian densities $L_i$:
\begin{equation}\label{Lagrangians2}
L_2=G_2(\phi, X), \ \  L_3=-G_3(\phi)\Box\phi, \ \ L_4=G_4(\phi)R, \ \ L_5=0,
\end{equation}
and the function $G_2(\phi, X)$ includes only zero and linear contributions of $X$.

Note that according to the GW170817 and GRB 170817A data $G_{4X} = 0$ and $G_5 = constant$ \citep{ez, Baker} and the discussed  subclass of Horndeski gravity is consistent with these constraints.

\subsection{Matter action}

Different modified gravitational models predict violations of equivalence principles which are basic for GR \citep{casola}. In scalar-tensor theories, the inertial mass and internal structure of a self-gravitating body depend on the local value of the scalar field. As a result, the laws of a self-gravitating body's motion depend on its internal structure and the GWEP violates \citep{casola}.  \citet{eardley} first considered the interaction of two point-like masses in the scalar-tensor theory and showed that in this case the influence of the scalar field on the internal structure of the body can be expressed through the assumption that the mass of the body is an arbitrary function of the scalar field. So the matter action for a system of point-like masses can be written as
\begin{equation}\label{matter_action}
S_m=-c^2\sum_a\int m_a(\phi)d\tau_a,
\end{equation}
where $m_a(\phi)$ are inertial masses of particles labeled by $a$ and $\tau_a$ is the proper time of the particle $a$ measured along its world-line $x^{\mu}_a$. From (\ref{matter_action}) it is clear that the mass $m_a(\phi)$ is position-dependent (because $\phi$ depends upon position) and hence the GWEP is violated. The stress-energy tensor of such action (\ref{matter_action}) and its trace take the forms
\begin{eqnarray}\label{stress-energy}
T^{\mu\nu}&=& \frac{c}{\sqrt{-g}}\sum_a m_a(\phi)\frac{u^{\mu}u^{\nu}}{u^0}\delta^3(\mathbfit{r}-\mathbfit{r}_a(t)), \nonumber\\
T&\equiv& g_{\mu\nu}T^{\mu\nu}=-\frac{c^3}{\sqrt{-g}}\sum_a \frac{m_a(\phi)}{u^0}\delta^3(\mathbfit{r}-\mathbfit{r}_a(t)),
\end{eqnarray}
where $u^{\mu}=d x^{\mu}_a/d \tau_a$ is four-velocity of the $a$-th particle, $d\tau=\sqrt{-ds^2}/c$, $ds^2=g_{\mu\nu}dx^{\mu}dx^{\nu}$ is an interval, $u_{\mu}u^{\mu}=-c^2$, and $\delta^3(\mathbfit{r}-\mathbfit{r}_a(t))$ is the three-dimensional Dirac delta function.

\section{Field equations in the weak-field limit} 
\label{sec:Field equations in the weak-field limit}
The purpose of our work is to study the gravitational radiation of binary pulsars. A pulsar is a strongly magnetized neutron star. The surface gravitational potential of such object $\Phi_{NS}=(G_NM_{NS})/(c^2R_{NS})=0.2$ (where $G_N$ is Newtonian gravitational constant, $M_{NS}$ is the mass of the neutron star, $R_{NS}$ is its radius). Therefore, the gravitational field around neutron star is very strong. However we are intend to calculate the energy flux carried away by gravitational radiation from  binary pulsars at  large distances from the source (e.g. at the position of the detector). At such large distances, the value of the surface potential of the source does not have a significant influence on the metric $g_{\mu\nu}$ and the scalar field $\phi$. Therefore, we can consider the perturbed field equations in a Minkowskian background \citep{will1, will11, will12}. Hence one expands the scalar and tensor fields in the limit of small velocities $(v/c\ll1)$:
\begin{eqnarray}\label{perturbations} 
\phi&=&\phi_0+\varphi, \nonumber\\
g_{\mu\nu}&=& \eta_{\mu\nu} + h_{\mu\nu}, 
\end{eqnarray} 
where $\eta_{\mu\nu}$ is the Minkowski background, $h_{\mu\nu}$ and $\varphi$ are the small perturbations of  tensor and scalar fields of order $O(v^2/c^2)$, respectively, $\phi_0$ is the asymptotic constant value of the scalar field far away from the source system (determined by the cosmological background solution).  Note that, in this paper we do not consider the effect of the cosmological evolution of the scalar field. The interesting aspects of time-dependent scalar field background were investigated earlier by \citet{babichev, brax, Sakstein, kopeikin, Arnoulx}. Also here we take a quasi-Minkowskian coordinate system. Such approximation of weak field limit and small velocities ($v/c \ll 1$) is the post-Newtonian (PN) expansion \citep{massivebd}.

Taking into account expressions~(\ref{perturbations}) and the fact that we consider case of Horndeski gravity without screening, the arbitrary functions $G_i(\phi, X)$ can be expanded in Taylor's series around the scalar asymptotic value:
\begin{eqnarray}\label{lag_func} 
G(\phi, X)&=&\sum_{m,n=0}^{\infty}G_{(m,n)}\varphi^mX^n,\ \ \ n\leq1,\nonumber\\
G_{(m,n)}&=&\frac{1}{m!n!}\ \frac{\partial^{m+n}}{\partial^m\phi\partial^nX}G(\phi, X)\bigg|_{\phi=\phi_0,X=0},\nonumber\\
G(\phi, X)&\equiv&G(\phi) \ \ \ \text{for} \ \ \ n=0.
\end{eqnarray} 
Here $G_{(m,n)}$ are constants.

In matter action (\ref{matter_action}) the inertial mass also is an arbitrary function of the scalar field, which can be expanded in Taylor's series around $\phi_0$ too:
\begin{equation}\label{inertial_mass} 
m_a(\phi)=m_a(\phi_0)\biggl[1+s_a\frac{\varphi}{\phi_0}-\frac{1}{2}\biggl(\frac{\varphi}{\phi_0}\biggr)^2(s_a-s_a^2-s'_a)+O(\varphi^3)\biggr].
\end{equation} 
Further we denote $m_a(\phi_0)$ as $m_a$ which is the inertial mass at the scalar asymptotic value. The quantities $s_a$ and $s_a'$ are the "first and second sensitivities". These parameters firstly were introduced by \citet{eardley}:
\begin{equation}\label{sensitivities} 
s_a\equiv \frac{\partial(\ln m_a)}{\partial(\ln \phi)}\bigg|_{\phi_0},\ \ \ s_a'\equiv \frac{\partial^2(\ln m_a)}{\partial(\ln \phi)^2}\bigg|_{\phi_0}.
\end{equation}

	Now we proceed directly to the obtaining of the field equations in the weak-field limit (in the general form for the Horndeski gravity they were presented by \citet{kobayashi, gao}). For $\phi$ we have:
\begin{eqnarray}\label{phi_field_equations}
&-&G_{2(0,1)}\Box\varphi+G_{2(0,1)}(\partial_\rho h^{\tau\rho}-\frac{1}{2}\eta^{\tau\alpha}\partial_\alpha h)\partial_\tau\varphi+2G_{3(1,0)}\Box\varphi \nonumber\\
&+&G_{2(0,1)}h_{\mu\nu}\partial^\mu\partial^\nu\varphi-G_{2(1,1)}\varphi\Box\varphi-G_{2(1,1)}\partial_\rho\varphi\partial^\rho\varphi\nonumber\\
&-&2G_{3(1,0)}h_{\mu\nu}\partial^\mu\partial^\nu\varphi-2G_{3(1,0)}(\partial_\rho h^{\tau\rho}-\frac{1}{2}\eta^{\tau\alpha}\partial_\alpha h)\partial_\tau\varphi\nonumber\\
&+&2G_{3(2,0)}\varphi\Box\varphi+2G_{3(2,0)}\partial_\rho\varphi\partial^\rho\varphi=\frac{16\pi}{c^4}\frac{\partial T}{\partial \varphi}+G_{2(1,0)} \nonumber\\
&+&G_{2(2,0)}\varphi+\frac{1}{2}G_{2(3,0)}\varphi^2-\biggl(\frac{1}{2}G_{2(1,1)}-2G_{3(2,0)}\biggr)\partial_{\rho}\varphi\partial^{\rho}\varphi\nonumber\\
&+&(G_{4(1,0)}+2G_{4(2,0)}\varphi)(\partial_\mu\partial_\nu h^{\mu\nu}-\Box h)+G_{4(1,0)}R[h^2].
\end{eqnarray} 
And for $g_{\mu\nu}$:
\begin{eqnarray}\label{g_field_equations} 
&-&\frac{1}{2}G_{2(0,0)}\eta_{\mu\nu}-\frac{1}{2}G_{2(0,0)}h_{\mu\nu}-\frac{1}{2}G_{2(1,0)}\eta_{\mu\nu}\varphi\nonumber\\
&-&\frac{1}{2}G_{2(1,0)}h_{\mu\nu}\varphi-\frac{1}{4}G_{2(2,0)}\eta_{\mu\nu}\varphi^2+\frac{1}{4}G_{2(0,1)}\eta_{\mu\nu}\partial_{\rho}\varphi\partial^{\rho}\varphi\nonumber\\
&-&\frac{1}{2}G_{2(0,1)}\partial_\mu\varphi\partial_\nu\varphi+G_{3(1,0)}\partial_\mu\varphi\partial_\nu\varphi-\frac{1}{2}G_{3(1,0)}\eta_{\mu\nu}\partial_{\rho}\varphi\partial^{\rho}\varphi\nonumber\\
&+&G_{4(0,0)}\biggl(\partial_\alpha \partial_\nu h^\alpha_\mu-\frac{1}{2}\Box h_{\mu\nu}-\frac{1}{2}\partial_\mu\partial_\nu h-\frac{1}{2}\eta_{\mu\nu} \partial_\alpha\partial_\beta h^{\alpha\beta}\nonumber\\
&+&\frac{1}{2}\eta_{\mu\nu} \Box h\biggr)+G_{4(1,0)}\eta_{\mu\nu}\Box\varphi-G_{4(1,0)}\partial_\mu\partial_\nu\varphi\nonumber\\
&+&G_{4(1,0)}\varphi\biggl(\partial_\alpha \partial_\nu h^\alpha_\mu-\frac{1}{2}\Box h_{\mu\nu}-\frac{1}{2}\partial_\mu\partial_\nu h-\frac{1}{2}\eta_{\mu\nu} \partial_\alpha\partial_\beta h^{\alpha\beta}\nonumber\\
&+&\frac{1}{2}\eta_{\mu\nu} \Box h\biggr)+G_{4(0,0)}G_{\mu\nu}[h^2]+G_{4(2,0)}\eta_{\mu\nu}\varphi\Box\varphi\nonumber\\
&+&G_{4(1,0)}h_{\mu\nu}\Box\varphi+\frac{1}{2}G_{4(1,0)}\partial_\rho\varphi(\partial_\nu h^\rho_{\mu}+\partial_\mu h^\rho_{\nu}-\eta^{\rho\alpha}\partial_\alpha h_{\mu\nu})\nonumber\\
&-&G_{4(1,0)}\eta_{\mu\nu}(\partial_\rho h^{\tau\rho}-\frac{1}{2}\eta^{\tau\alpha}\partial_\alpha h)\partial_\tau\varphi-G_{4(2,0)}\varphi\partial_\mu\partial_\nu\varphi\nonumber\\
&-&G_{4(1,0)}\eta_{\mu\nu}h_{\rho\sigma}\partial^\rho\partial^\sigma\varphi+G_{4(2,0)}\eta_{\mu\nu}\partial_{\rho}\varphi\partial^{\rho}\varphi\nonumber\\
&-&G_{4(2,0)}\partial_\mu\varphi\partial_\nu\varphi=\frac{8\pi}{c^4}T_{\mu\nu},
\end{eqnarray} 
where $G_{\mu\nu}[h^2]$ is the part of Einstein tensor of order $h_{\mu\nu}h^{\mu\nu}$.
	
	Taking into account (\ref{perturbations}) we obtain the expressions for stress-energy tensor (\ref{stress-energy}), its trace, and $\frac{\partial T}{\partial \varphi}$ in the near zone:
	\begin{eqnarray}\label{stress-energy_1}
	T^{\mu\nu}=& &\sum_a m_au^{\mu}u^{\nu}\biggl(1-\frac{h^k_k}{2}-\frac{v_a^2}{2c^2}+s_a\frac{\varphi}{\phi_0}\biggr)\delta^3(\mathbfit{r}-\mathbfit{r}_a(t)), \nonumber\\
	T=&-&c^2\sum_a m_a\biggl(1-\frac{h^k_k}{2}-\frac{v_a^2}{2c^2}+s_a\frac{\varphi}{\phi_0}\biggr)\delta^3(\mathbfit{r}-\mathbfit{r}_a(t)), \nonumber\\
	\frac{\partial T}{\partial \varphi}=&-&c^2\sum_a \frac{m_a}{\phi_0}\biggl[s_a\biggl(1-\frac{h^k_k}{2}-\frac{v_a^2}{2c^2}\biggr)-(s_a-s_a^2-s'_a)\frac{\varphi}{\phi_0}\biggr]\nonumber\\
&\times&\delta^3(\mathbfit{r}-\mathbfit{r}_a(t)),
	\end{eqnarray}
	where $v_a$ is the velocity of the object labeled in $a$.
	
	The terms $G_{2(0,0)}$ and $G_{2(1,0)}$ are responsible for effects of DE. \citet{ashtekar} show that such effects on gravitational waves from isolated systems are insignificant, so we can neglect these terms.
	
	\section{Post-Newtonian solutions}\label{sec:Post-newtonian solutions}
	
	Before  investigating the model in the far zone and studying gravitational radiation at the point of the detector, we must solve the field equations in the near zone, where the gravitational radiation is generated.
	
	The field equations~(\ref{phi_field_equations}) and (\ref{g_field_equations}) within the post-Newtonian (PN) approximation in the 1st PN order $O(v/c)^4$ take the forms \citep{hohmann}:
	\begin{eqnarray}\label{phi_1} 
	&-&(G_{2(0,1)}-2G_{3(1,0)})\Box\varphi-G_{2(2,0)}\varphi+G_{4(1,0)}(\Box h-\partial_\mu\partial_\nu h^{\mu\nu})\nonumber\\
&&=\frac{16\pi}{c^4}\frac{\partial T}{\partial \varphi},\nonumber\\
	&&G_{4(0,0)}\biggl(\partial_\alpha \partial_\nu h^\alpha_\mu-\frac{1}{2}\Box h_{\mu\nu}-\frac{1}{2}\partial_\mu\partial_\nu h -\frac{1}{2}\eta_{\mu\nu} \partial_\alpha\partial_\beta h^{\alpha\beta} \nonumber\\
	&+&\frac{1}{2}\eta_{\mu\nu} \Box h\biggr)+G_{4(1,0)}\eta_{\mu\nu}\Box\varphi-G_{4(1,0)}\partial_\mu\partial_\nu\varphi=\frac{8\pi}{c^4}T_{\mu\nu}.
	\end{eqnarray}

	Further, we introduce the following notations:
	\begin{eqnarray}\label{theta} 
	\theta_{\mu\nu}&=&h_{\mu\nu}-\frac{1}{2}\eta_{\mu\nu}h-\frac{G_{4(1,0)}}{G_{4(0,0)}}\eta_{\mu\nu}\varphi, \nonumber\\
	\theta&=&-h-4\frac{G_{4(1,0)}}{G_{4(0,0)}}\varphi .
	\end{eqnarray} 
	
	The choice of the transverse gauge $\partial_\mu \theta^{\mu\nu}=0$ reduces the field equations as follows:
	\begin{equation}\label{g_1} 
	\Box\theta_{\mu\nu}=-\frac{16\pi}{c^4G_{4(0,0)}}T_{\mu\nu},
	\end{equation} 
	\begin{equation}\label{phi_1} 
	\Box\varphi-m_\varphi^2\varphi=\frac{16\pi}{c^4}c_\varphi S,
	\end{equation} 
where
\begin{equation}\label{ms} 
m_\varphi^2=\frac{G_{2(2,0)}}{2G_{3(1,0)}-G_{2(0,1)}-3\frac{G_{4(1,0)}^2}{G_{4(0,0)}}},\\
\end{equation} 
\begin{equation}\label{cphi} 
c_\varphi=-\frac{G_{4(1,0)}}{2G_{4(0,0)}\biggl(2G_{3(1,0)}-G_{2(0,1)}-3\frac{G_{4(1,0)}^2}{G_{4(0,0)}}\biggr)},\\
\end{equation}
\begin{equation}\label{s} 
S=T-\frac{2G_{4(0,0)}}{G_{4(1,0)}}\frac{\partial T}{\partial \varphi}.
\end{equation} 
The equation~(\ref{phi_1}) is an analogue of inhomogeneous Klein-Gordon one, where the parameter $m_\varphi$ is the inverse Compton wavelength of the scalar field. In this paper we work in the CGS system, thus here and further the scalar field mass $m_\varphi$ has the dimension of inverse length [cm$^{-1}$].

The next step is obtaining the leading order of the static solution for the scalar field. Using (\ref{phi_1}), (\ref{stress-energy_1}) and (\ref{s}) it is possible to derive the following expression \citep{hohmann}:
\begin{equation}\label{phi} 
\varphi=\frac{4c_\varphi}{c^2}\sum_a\frac{m_a}{r_a}\biggl(1-\frac{2s_a}{\phi_0}\frac{G_{4(0,0)}}{G_{4(1,0)}}\biggr)e^{-m_\varphi r_a},
\end{equation} 
where $r_a=|\mathbfit{r}-\mathbfit{r}_a(t)|$.

Further, the solution of the tensor field equations~(\ref{g_1}) within the 1PN approximation in the near zone is 
\begin{eqnarray}\label{h} 
\theta_{00}&=&\frac{4}{c^2G_{4(0,0)}}\sum_a\frac{m_a}{r_a}+O\biggl(\frac{v}{c}\biggr)^4, \nonumber\\
\theta_{ij}&=&\frac{4v_iv_j}{c^4G_{4(0,0)}}\sum_a\frac{m_a}{r_a}+O\biggl(\frac{v}{c}\biggr)^6, \nonumber\\
\theta&=&-\frac{4}{c^2G_{4(0,0)}}\sum_a\frac{m_a}{r_a}+O\biggl(\frac{v}{c}\biggr)^4.
\end{eqnarray} 
Thus, the leading order of metric perturbation is defined as \citep{hohmann}:
\begin{eqnarray}\label{h} 
h_{00}&=&\frac{2}{c^2G_{4(0,0)}}\sum_a\frac{m_a}{r_a}+\frac{4c_\varphi}{c^2}\frac{G_{4(1,0)}}{G_{4(0,0)}} \sum_a\frac{m_a}{r_a}e^{-m_\varphi r_a}\nonumber\\
&&\times\biggl(1-\frac{2s_a}{\phi_0}\frac{G_{4(0,0)}}{G_{4(1,0)}}\biggr)+O\biggl(\frac{v}{c}\biggr)^4,\nonumber\\
h_{ij}&=&\delta_{ij}\biggl[\frac{2}{c^2G_{4(0,0)}}\sum_a\frac{m_a}{r_a}-\frac{4c_\varphi}{c^2}\frac{G_{4(1,0)}}{G_{4(0,0)}} \sum_a\frac{m_a}{r_a}e^{-m_\varphi r_a}\nonumber\\
&&\times\biggl(1-\frac{2s_a}{\phi_0}\frac{G_{4(0,0)}}{G_{4(1,0)}}\biggr)\biggr]+O\biggl(\frac{v}{c}\biggr)^4, \nonumber\\
h&=&\frac{4}{c^2G_{4(0,0)}}\sum_a\frac{m_a}{r_a}-\frac{16c_\varphi}{c^2}\frac{G_{4(1,0)}}{G_{4(0,0)}}\sum_a\frac{m_a}{r_a}e^{-m_\varphi r_a} \nonumber\\
&&\times\biggl(1-\frac{2s_a}{\phi_0}\frac{G_{4(0,0)}}{G_{4(1,0)}}\biggr)+O\biggl(\frac{v}{c}\biggr)^4,
\end{eqnarray} 
with $h_{oi}=O(v/c)^3$. Here $\delta_{ij}$ is the Kronecker delta.

\section{EIH equations of motion}\label{sec:EIH equations of motion}

The gravitational weak equivalence principle states  \citep{casola}: 
\begin{quote}
	Test particles behave, in a gravitational field and in vacuum, independently of their properties. \end{quote}
The GWEP is one of the principles, which works in GR but can be violated in alternative theories of gravity \citep{casola}. In particular, in the scalar-tensor models the dependence of the inertial mass upon the sensitivity leads to the violation of GWEP. Sensitivity shows the changing of a compact object's mass as it moves relatively to the additional field. Therefore, different bodies react not in the same manner to the motion relative to the ambient field. Thus, they move along different trajectories. Due to the violation of GWEP the conservative orbital dynamics of compact systems modifies. To find the explicit form of the sensitivity influence on the equations of motion, we use the method suggested by \citet{eih}.

The equations of motion for the mass $m_a$ can be obtained from the matter Lagrangian:
\begin{eqnarray}\label{eih} 
L_{EIH}=&-&c^2\sum_a\int m_a(\phi)\frac{d\tau_a}{dt}\nonumber\\
=&-&c^2\sum_a m_a(\phi)\sqrt{-g_{00}-2g_{0i}\frac{v^i_a}{c}-g_{ij}\frac{v^i_av^j_a}{c^2}}\nonumber\\
=&-&\sum_a m_ac^2\biggl\{1-\frac{v^2_a}{2c^2}-\sum_{b\neq a}\biggl[\frac{1}{c^2G_{4(0,0)}}\frac{m_b}{r_{ab}}\nonumber\\
&+&\frac{2c_\varphi}{c^2}\frac{G_{4(1,0)}}{G_{4(0,0)}}\frac{m_b}{r_{ab}}\biggl(1-\frac{2s_b}{\phi_0}\frac{G_{4(0,0)}}{G_{4(1,0)}}\biggr)e^{-m_\varphi r_{ab}}\\
&-&\frac{4s_a c_\varphi}{c^2\phi_0}\frac{m_b}{r_{ab}}\biggl(1-\frac{2s_b}{\phi_0}\frac{G_{4(0,0)}}{G_{4(1,0)}}\biggr)e^{-m_\varphi r_{ab}}\biggr]+O\biggl(\frac{v}{c}\biggr)^4\biggr\}\nonumber,
\end{eqnarray} 
there $r_{ab}=|\mathbfit{r}_a(t)-\mathbfit{r}_b(t)|$. From the equation of motion (\ref{eih}) we can identify the effective gravitational "constant":
\begin{eqnarray}\label{gab} 
G_{ab}&=&\biggl[\frac{1}{G_{4(0,0)}}+2c_\varphi\frac{G_{4(1,0)}}{G_{4(0,0)}}e^{-m_\varphi r_{ab}}\biggl(1-\frac{2s_b}{\phi_0}\frac{G_{4(0,0)}}{G_{4(1,0)}}\biggr)\nonumber\\
&&-\frac{4s_a c_\varphi}{\phi_0}\biggl(1-\frac{2s_b}{\phi_0}\frac{G_{4(0,0)}}{G_{4(1,0)}}\biggr)e^{-m_\varphi r_{ab}}\biggr].
\end{eqnarray} 
This result is symmetric under interchange of all particle pairs \citep{hou}.

The corresponding n-body equations of motion up to Newtonian order are defined as follows
\begin{eqnarray}\label{aa} 
\mathbfit{a}_a=-\sum_{a\neq b}\frac{\mathcal{G}_{ab}m_b}{r_{ab}^2}\hat{\mathbfit{r}}_{ab},
\end{eqnarray} 
with
\begin{eqnarray}\label{geff} 
\mathcal{G}_{ab}&=&\frac{1}{G_{4(0,0)}}\biggl\{1+(1+m_\varphi r_{ab})e^{-m_\varphi r_{ab}}\biggl[2c_\varphi G_{4(1,0)}\\
&&\times\biggl(1-\frac{2s_b}{\phi_0}\frac{G_{4(0,0)}}{G_{4(1,0)}}\biggr)-\frac{4s_a c_\varphi G_{4(0,0)}}{\phi_0}\biggl(1-\frac{2s_b}{\phi_0}\frac{G_{4(0,0)}}{G_{4(1,0)}}\biggr)\biggr]\biggr\}\nonumber,
\end{eqnarray} 
where $\mathbfit{a}_a$ is the acceleration of the $a$-th object, $\hat{\mathbfit{r}}_{ab}$ is the unit direction vector from the $b$-th object to the $a$-th one. The scalar field mass is responsible for DE effect. Therefore, the manifestations of influence of this effect start from the distances much larger than the distance between components in binary pulsars. Thus we use the approximation $m_\varphi r_{ab}\ll1$ and $e^{-m_\varphi r_{ab}}\to 1$. In this case, the effective gravitational constant between components in binary pulsars takes the following form:
\begin{eqnarray}\label{geff_1} 
\mathcal{G}_{ab}&=&\frac{1}{G_{4(0,0)}}\biggl[1+2c_\varphi G_{4(1,0)}\biggl(1-\frac{2s_b}{\phi_0}\frac{G_{4(0,0)}}{G_{4(1,0)}}\biggr)\nonumber\\
&&-\frac{4s_a c_\varphi G_{4(0,0)}}{\phi_0}\biggl(1-\frac{2s_b}{\phi_0}\frac{G_{4(0,0)}}{G_{4(1,0)}}\biggr)\biggr].
\end{eqnarray}

Now let us consider an orbital dynamics of a binary system with compact objects. A motion in binary system obeys the Kepler's third law:
\begin{equation}\label{Kepler} 
a^3(2\pi/P_b)^2=\mathcal{G}_{12}m
\end{equation} 
and the orbital binding energy of such system is
\begin{equation}\label{energy} 
E=-\frac{\mathcal{G}_{12}m\mu}{2a},
\end{equation} 
here $a$ is the semi-major axis, $\mathcal{G}_{12}$ is the effective gravitational coupling constant between two compact objects, $m=m_1+m_2$, and $\mu=m_1m_2/m$, $P_b$ is the orbital period of a binary system.

The most significant dissipative effect is the orbital period decay due to the emission of gravitational radiation. The energy loss can be expressed via the first derivative of the orbital period using equations~(\ref{Kepler}) and (\ref{energy}):
\begin{equation}\label{energy_losses} 
\frac{\dot E}{E}=-\frac{2}{3}\frac{\dot P_b}{P_b}.
\end{equation} 
Thus, we find that the orbital decay of the binary pulsars is directly determined by the energy loss of the system.

\section{Graviational radiation from binary pulsars}\label{sec:GRAVITATIONAL RADIATION FROM COMPACT BINARIES}

In this section, we focus on the dissipative effects, calculate the rate of the energy loss due to the emission of gravitational radiations (including monopole, dipole, quadrupole, and dipole-octupole radiations), and derive their contributions to the change of the orbital period. Nevertheless, before we move to direct computations, we need to obtain the stress-energy pseudotensor.

\subsection{Effective stress-energy pseudotensor}

Far away from the local system the stress-energy tensor  of the source $T_{\mu\nu}$ vanishes. However, the influence of the local system on the flat space-time remains in the form of gravitational  radiation. This radiation has energy and momentum and is described by effective stress-energy pseudotensor.

There are many different methods to define the energy-momentum pseudotensor \citep{petrov, saffer}. Authors of the last work investigate four ones: the second variation of the action under short-wavelength averaging, the second perturbation of the field equations in the short-wavelength approximation, the construction of an energy complex leading to a Landau-Lifshitz tensor, and the using of Noether's theorem in field theories about a flat background. All these ways lead to different results but yield the same rate of energy loss. We apply the Noether current method. It is suitable for our purposes and the fact that the method yields a not symmetric stress-energy pseudotensor is not significant. However, the discussed approach has the serious problems with definition of angular momentum of an isolated gravitating system. Fortunately, in our work, the angular momentum is not required and chosen method provides the correct result.

First, we consider a general action:
\begin{eqnarray}\label{action_noether} 
S&=&\int d^4x\sqrt{-g}[L_{g}(g_{\mu\nu}, \partial_\alpha g_{\mu\nu}, \partial_\alpha \partial_\beta g_{\mu\nu}, \phi, \partial_\alpha\phi, \partial_\alpha \partial_\beta \phi) \nonumber\\
&&+L_{m}(q, \partial_\alpha q, \partial_\alpha \partial_\beta q, \phi, \partial_\alpha\phi, \partial_\alpha \partial_\beta \phi)],
\end{eqnarray} 
where $L_{g}$ is a gravitational Lagrangian density, $L_{m}$ is a matter Lagrangian density, $q$ is the matter fields. 

Far away from the local system the conservation laws take the following form:
\begin{equation}\label{conservation law} 
\partial_\alpha(\sqrt{-g}[T^\alpha_\gamma+t^\alpha_\gamma])=0.
\end{equation}

The part with $L_m$ in (\ref{action_noether}) gives the canonical matter stress-energy tensor which is equivalent to Hilbert stress-energy tensor $T_{\mu\nu}$. The remaining quantity $t^\alpha_\gamma$ is our sought-for pseudotensor. The Noether current method defines a pseudotensor as:
\begin{eqnarray}\label{pseudotensor} 
t^\alpha_\gamma=&-&\frac{\partial L_{g}}{\partial (\partial_\alpha g_{\mu\nu})}\partial_\gamma g_{\mu\nu}+\partial_\beta\biggl(\frac{\partial L_{g}}{\partial (\partial_\alpha\partial_\beta g_{\mu\nu})}\biggr)\partial_\gamma g_{\mu\nu}\nonumber\\
&-&\frac{\partial L_{g}}{\partial (\partial_\alpha\partial_\beta g_{\mu\nu})}\partial_\beta\partial_\gamma g_{\mu\nu}-\frac{\partial L_{g}}{\partial (\partial_\alpha \phi)}\partial_\gamma\phi\nonumber\\
&-&\frac{\partial L_{g}}{\partial (\partial_\alpha\partial_\beta \phi)}\partial_\beta\partial_\gamma \phi+R\partial_\beta\biggl(\frac{\partial L_{g}}{\partial (\partial_\alpha\partial_\beta \phi)}\biggr)\partial_\gamma \phi+\delta^\alpha_\gamma L_{g}.\nonumber\\
\end{eqnarray} 

Now we return to the considering subclass of Horndeski theory. The gravitational part of Lagrangian density is given by the expressions (\ref{Lagrangians2}). Using the expansions (\ref{perturbations}) the Lagrangian densities are reduced to
\begin{eqnarray}\label{Lagrangians_2}
L_2&=&\frac{c^4}{16\pi}\biggl(G_{2(2, 0)}\varphi^2-\frac{1}{2}G_{2(0, 1)}\partial_\mu\varphi\partial^\mu\varphi\biggr),\nonumber\\
L_3&=&\frac{c^4}{16\pi}G_{3(1, 0)}\partial_\mu\varphi\partial^\mu\varphi, \nonumber\\
L_4&=&\frac{c^4}{16\pi}\biggl[\frac{G_{4(0, 0)}}{4}\biggl(4\partial_\mu\partial_\nu h^{\mu\nu}-4\Box h-8h^{\mu\nu}\partial_\mu\partial_\alpha h^\alpha_\nu \nonumber\\
&&+4h^{\mu\nu}\Box h_{\mu\nu}-4\partial_\alpha h^{\alpha\mu}\partial_\nu h^\nu_\mu+4\partial_\mu h^{\mu\nu}\partial_\nu h +4h^{\mu\nu}\partial_\mu\partial_\nu h\nonumber\\
&&+3\partial_\alpha h^{\mu\nu}\partial^\alpha h_{\mu\nu}-\partial_\mu h\partial^\mu h-2\partial_\alpha h^{\mu\nu}\partial_\mu h^\alpha_\nu\biggr)\nonumber\\
&&+G_{4(1, 0)}\varphi(\partial_\mu\partial_\nu h^{\mu\nu}-\Box h)\biggr],\nonumber\\
L_5&=&0.
\end{eqnarray}

According to the four-dimensional analogue of the  Ostro\-gradskii-Gauss theorem  we can throw out the total derivatives from the action (\ref{act}). The remaining part is 
\begin{eqnarray}\label{Lagrangians_3}
\sqrt{-g} L_2&=&\sqrt{-g}\frac{c^4}{16\pi}\biggl(G_{2(2, 0)}\varphi^2-\frac{1}{2}G_{2(0, 1)}\partial_\mu\varphi\partial^\mu\varphi\biggr),\nonumber\\
\sqrt{-g} L_3&=&\sqrt{-g}\frac{c^4}{16\pi}\biggl(G_{3(1, 0)}\partial_\mu\varphi\partial^\mu\varphi\biggr), \nonumber\\
\sqrt{-g}L_4&=&\sqrt{-g}\frac{c^4}{16\pi}\biggl[\frac{G_{4(0, 0)}}{4}\biggl(2\partial_\alpha h^{\alpha\mu}\partial_\nu h^\nu_\mu-2\partial_\mu h^{\mu\nu}\partial_\nu h\nonumber\\
&&-\partial_\alpha h^{\mu\nu}\partial^\alpha h_{\mu\nu}+\partial_\mu h\partial^\mu h\biggr)\nonumber\\
&&-G_{4(1, 0)}\partial_\mu\varphi(\partial_\nu h^{\mu\nu}-\partial^\mu h)\biggr],\nonumber\\
\sqrt{-g} L_5&=&0.
\end{eqnarray}

Now we express equation~(\ref{pseudotensor}) in respect the equation~(\ref{Lagrangians_3}):
\begin{eqnarray}\label{pseudotensor_2} 
t^\alpha_\gamma=&&\frac{c^4}{16\pi}\biggl\{G_{2(0, 1)}\partial_\gamma\varphi\partial^\alpha\varphi-2G_{3(1,0)}\partial_\gamma\varphi\partial^\alpha\varphi+\frac{G_{4(0, 0)}}{4}\nonumber\\
&\times&\biggl(-4\partial_\nu h^{\nu\mu}\partial_\gamma h^\alpha_\mu+2\partial_\mu h^{\mu\alpha}\partial_\gamma h+2\partial_\gamma h^{\mu\nu}\partial^\alpha h_{\mu\nu}\nonumber\\
&-&2\partial_\gamma h\partial^\alpha h+2\partial_\gamma h^{\alpha\nu}\partial_\nu h\biggr)+G_{4(1, 0)}\partial_\mu\varphi \partial_\gamma h^{\mu\alpha}\nonumber\\
&+&G_{4(1, 0)}\partial_\gamma\varphi(\partial_\nu h^{\alpha\nu}-\partial^\alpha h)-G_{4(1, 0)}\partial^\alpha\varphi\partial_\gamma h\nonumber\\
&+&\delta^\alpha_\gamma\biggl[G_{2(2, 0)}\varphi^2-\frac{1}{2}G_{2(0, 1)}\partial_\mu\varphi\partial^\mu\varphi+G_{3(1,0)}\partial_\mu\varphi\partial^\mu\varphi\nonumber\\
&+&\frac{G_{4(0, 0)}}{4}\biggl(2\partial_\alpha h^{\alpha\mu}\partial_\nu h^\nu_\mu-2\partial_\mu h^{\mu\nu}\partial_\nu h-\partial_\alpha h^{\mu\nu}\partial^\alpha h_{\mu\nu}\nonumber\\
&+&\partial_\mu h\partial^\mu h\biggr)-G_{4(1, 0)}\partial_\mu\varphi(\partial_\nu h^{\mu\nu}-\partial^\mu h)\biggr]\biggr\}.
\end{eqnarray} 

Further we turn to the new variables (\ref{theta}) and impose transverse-traceless (TT) gauge  including two conditions $\partial_\mu\theta^{\mu\nu}=0 \text{ and } \eta_{\mu\nu}\theta^{\mu\nu}=0$. Finally we obtain
\begin{eqnarray}\label{pseudotensor_3} 
t^\alpha_\gamma=&&\frac{c^4}{16\pi}\biggl\{\frac{G_{4(0, 0)}}{2}\ \partial_\gamma \theta^{\mu\nu}\partial^\alpha \theta_{\mu\nu}-\partial_\gamma\varphi\partial^\alpha\varphi\biggl(2G_{3(1,0)}\nonumber\\
&-&G_{2(0, 1)}-3\frac{G^2_{4(1,0)}}{G_{4(0,0)}}\biggr)+\delta^\alpha_\gamma\biggl[G_{2(2, 0)}\varphi^2\nonumber\\
&-&\frac{G_{4(0, 0)}}{4}\partial_\alpha \theta^{\mu\nu}\partial^\alpha \theta_{\mu\nu}+\frac{1}{2}\partial_\mu\varphi\partial^\mu\varphi\biggl(2G_{3(1,0)}-G_{2(0, 1)}\nonumber\\
&-&3\frac{G^2_{4(1,0)}}{G_{4(0,0)}}\biggr)\biggr]\biggr\}=\frac{c^4}{16\pi}\biggl[\frac{G_{4(0, 0)}}{2}\ \partial_\gamma \theta^{\mu\nu}\partial^\alpha \theta_{\mu\nu}\nonumber\\
&+&\frac{G_{4(1,0)}}{2G_{4(0,0)}c_\varphi}\partial_\gamma\varphi\partial^\alpha\varphi+\delta^\alpha_\gamma\biggl(G_{2(2, 0)}\varphi^2\nonumber\\
&-&\frac{G_{4(0, 0)}}{4}\partial_\alpha \theta^{\mu\nu}\partial^\alpha \theta_{\mu\nu}-\frac{G_{4(1,0)}}{4G_{4(0,0)}c_\varphi}\partial_\mu\varphi\partial^\mu\varphi\biggr)\biggr].
\end{eqnarray} 
In the last step  the expression for $c_\varphi$ from (\ref{cphi}) is used. Thus, the final form of Noether's pseudotensor in TT-gauge is derived.

Now we have everything necessary for the calculations of tensor and scalar energy fluxes.

\subsection{Tensor and scalar energy fluxes}

Gravitational waves carry energy and bend the space-time. Different momentum and energy characteristics (densities and fluxes) of the gravitational waves are expressed in terms of stress-energy pseudotensor. The component $t_{0i}$ is responsible for the energy flux \citep{will1, will11, will12}, thus the average rate of the binding energy change of binary system is defined as
\begin{equation}\label{flux} 
\left<\dot E\right> = -cr^2\int d\Omega \left<t_{TT}^{0r}\right>,
\end{equation} 
where the angular brackets represent a time average over a period of the system's motion, $\Omega$ is the solid angle, TT means the transverse-traceless gauge.

In GR, the energy flux appears only due to the propagation of tensor mode, but in the standard massive scalar-tensor theories, gravitational radiation comes from both scalar and tensor modes. In the wave zone (far zone) the matter is absent and $T^{\alpha\gamma}=0$, so the conservation law is $\partial_\alpha t_{TT}^{\alpha\gamma}=0$. Since there are no mixed components inside the pseudotensor $t^{\alpha\gamma}_{TT}$ ($\theta_{\mu\nu}$ and $\varphi$ are decoupled). The energy-momentum pseudotensors (i.e., Noether currents) of the tensor 
\begin{eqnarray}\label{tensor_pseudotensor} 
t_{\alpha\gamma}^{TT}(\theta^{TT}_{\mu\nu})&=&\frac{c^4}{16\pi}\biggl[\frac{G_{4(0, 0)}}{2}\ \partial_\gamma \theta_{TT}^{\mu\nu}\partial_\alpha \theta^{TT}_{\mu\nu}\nonumber\\
&&-\delta_{\alpha\gamma}\biggl(\frac{G_{4(0, 0)}}{4}\partial_\mu \theta_{TT}^{\mu\nu}\partial^\mu \theta^{TT}_{\mu\nu}\biggr)\biggr]
\end{eqnarray} 
and scalar gravitational waves
\begin{eqnarray}\label{scalar_pseudotensor} 
t^{TT}_{\alpha\gamma}(\varphi)&=&\frac{c^4}{16\pi}\biggl[\frac{G_{4(1,0)}}{2G_{4(0,0)}c_\varphi}\partial_\gamma\varphi\partial_\alpha\varphi\nonumber\\
&&+\delta_{\alpha\gamma}\biggl(G_{2(2, 0)}\varphi^2-\frac{G_{4(1,0)}}{4G_{4(0,0)}c_\varphi}\partial_\mu\varphi\partial^\mu\varphi\biggr)\biggr]\nonumber\\
\end{eqnarray} 
are respectively conserved and we can investigate them separately.

\subsubsection{Tensor energy flux}

According to (\ref{flux}) the average energy flux radiated in gravitational waves due to tensor part is
\begin{eqnarray}\label{gravitational flux} 
\left<\dot E_g\right> &=& -cr^2\int d\Omega \left<t^{0r}_{TT}(\theta_{\mu\nu})\right> \nonumber\\
&=&\frac{c^5r^2}{16\pi}\int d\Omega \ \left<\frac{G_{4(0, 0)}}{2}\ \partial_0 \theta_{TT}^{\mu\nu}\partial_r\theta^{TT}_{\mu\nu}\right>.
\end{eqnarray} 
The tensor mode is massless and propagates with the speed of light, $\theta_{ij}(t,r)$ takes the form $(1/r)f_{ij} (t- r/c)$ therefore at large distances $\partial_r\theta_{ij}=-\partial_0\theta_{ij}$ at the leading order. Using this fact, the tensor energy equation~(\ref{gravitational flux}) can be simplified to
\begin{equation}\label{gravitational flux1} 
\left<\dot E_g\right> =-\frac{c^5r^2G_{4(0, 0)}}{32\pi}\int d\Omega \ \left< \partial_0 \theta_{TT}^{\mu\nu}\partial_0\theta^{TT}_{\mu\nu}\right>.
\end{equation} 

Now we return to the equation~(\ref{g_1}). The formal solution is
\begin{equation}\label{formal solution} 
\theta_{\mu\nu} =\frac{4}{c^4G_{4(0, 0)}}\int_N d^3 \mathbfit{r}' \frac{T_{\mu\nu}(t-|\mathbfit{r}-\mathbfit{r}'|/c)}{|\mathbfit{r}-\mathbfit{r}'|}.
\end{equation} 
Here, source point $\mathbfit{r}'$ belongs to the near zone $N$, whereas the field point $\mathbfit{r}$ is located in the far zone (wave zone), such that $|\mathbfit{r}'|\ll |\mathbfit{r}|$. Taking into account this condition, we can expand the integrand in powers of $(\mathbfit{n} \times \mathbfit{r}')$ in the slow-motion approximation
\begin{equation}\label{formal solution_2} 
\theta_{\mu\nu} =\frac{4}{rc^4G_{4(0, 0)}}\sum_{l=0}^{\infty}\frac{1}{c^ll!}\frac{\partial^l}{\partial t^l}\int_N T_{\mu\nu}(t-r/c,\mathbfit{r}')(\mathbfit{n} \times \mathbfit{r}')^l d^3 \mathbfit{r}',
\end{equation} 
where $\mathbfit{n}= \mathbfit{r}/r$ is the unit vector in the $\mathbfit{r}$ direction. Using the conservation law $\partial_\nu T^{\mu\nu}=0$, we can express the spatial components $\theta_{ij}$ up to leading order ($l=0$) as
\begin{eqnarray}\label{spatial solution} 
\theta_{ij} &=&\frac{4}{rc^4G_{4(0, 0)}}\int T_{ij}(t-r/c,\mathbfit{r}') d^3 \mathbfit{r}'\nonumber\\
&=&\frac{2}{rc^6G_{4(0, 0)}}\frac{\partial^2}{\partial t^2}\int T_{00}(t-r/c,\mathbfit{r}')r_i'r_j' d^3 \mathbfit{r}'.
\end{eqnarray} 
There is only the quadrupole moment of $T_{00}$, like in GR. So the monopole and dipole contributions are absent in the tensor gravitational radiation because the tensor graviton is a massless spin-2 particle. 

The quantity $T_{00}$ is the energy density. The leading PN order contribution from $T_{00}$ is
\begin{equation}\label{t00} 
T_{00}=\sum_a m_ac^2\delta^3(\mathbfit{r}-\mathbfit{r}_a(t)).
\end{equation} 
Substituting this expression into equation~(\ref{spatial solution}) we obtain
\begin{equation}\label{spatial solution2} 
\theta_{ij} =\frac{2}{rc^4G_{4(0, 0)}}\frac{\partial^2}{\partial t^2}M_{ij}\bigg|_{ret},
\end{equation} 
where
\begin{equation}\label{quadrupole moment} 
M_{ij} =\sum_a m^a(\phi)r_{i}^a(t)r_{j}^a(t)
\end{equation} 
is the mass quadrupole moment. The subscript "ret" means that the quantity $M_{ij}$ is evaluated at the retarded time $t-r/c$ .

Now it is possible to find the average energy flux of tensor sector in terms of the mass quadrupole moments:
\begin{equation}\label{gravitational flux2} 
\left<\dot E_g\right> =-\frac{1}{5c^5G_{4(0, 0)}} \left< \dddot{M}^{kl}\dddot{M}_{kl}-\frac{1}{3}(\dddot{M}^{kk})^2\right>.
\end{equation} 
For the integration over the solid angle we use the fact that the $\theta_{ij}^{TT}=\Lambda_{ij,kl}\theta^{kl}$. Here the projector $\Lambda_{ij,kl}$ is the Lambda tensor (see \citet{massivebd}). The overdots represent derivatives with respect to coordinate time.

In our work, we consider only binary systems with quasi-circular orbits which can be parameterized by
\begin{eqnarray}\label{quasicircular} 
x_1(t)&=&-R_1\cos(\omega t), \ y_1(t)=-R_1\sin(\omega t), \ z_1=0;\nonumber\\
x_2(t)&=&R_2\cos(\omega t), \ y_2(t)=R_2\sin(\omega t),\ z_2=0,
\end{eqnarray} 
here by $R_a$ we denote the orbital radii of the binary system components, and $\omega$ is the orbital frequency. Using the Kepler's third law (\ref{Kepler}) we find the final form of the average energy flux radiated in gravitational waves due to tensor sector:
\begin{eqnarray}\label{gravitational flux3} 
\left<\dot E_g\right> =&-&\frac{32\mu^2(\mathcal{G}_{12}m)^3}{5c^5G_{4(0,0)}R^5}=-\frac{32\mu^2m^3}{5c^5G^4_{4(0,0)}R^5}\nonumber\\
&\times&\biggl\{1+\biggl[2c_\varphi G_{4(1,0)}\biggl(1-\frac{2s_b}{\phi_0}\frac{G_{4(0,0)}}{G_{4(1,0)}}\biggr)\nonumber\\
&-&\frac{4s_a c_\varphi G_{4(0,0)}}{\phi_0}\biggl(1-\frac{2s_b}{\phi_0}\frac{G_{4(0,0)}}{G_{4(1,0)}}\biggr)\biggr]\biggr\}^3,
\end{eqnarray} 
where $R=R_1+R_2$ and in the case of quasi-circular orbit in equation~(\ref{Kepler}) $R=a$. The quantity $\mathcal{G}_{12}$ is the effective gravitational constant between components of binary system (\ref{geff_1}).

\subsubsection{Scalar energy flux}

Now we obtain the change of the binding energy due to the scalar radiation. Our consideration starts from a formal solution of equation~(\ref{phi_1}). The obtaining of such type solution by using Green's function method is described in detail in papers of \citet{morse, massivebd, zhang}. We fully follow them  and present the final expression for the formal solution
\begin{eqnarray}\label{fi} 
\varphi=&-&\frac{4c_\varphi}{rc^4}\int_0^\infty dz J_1(z)\sum_{l=0}^\infty\frac{1}{c^ll!}\frac{\partial^l}{\partial t^l}\int_Nd^3\mathbfit{r}'(\mathbfit{n} \times \mathbfit{r}')^l\nonumber\\
&\times&\biggl[S(t-r/c,\mathbfit{r}')-\frac{S(t-ru(r,z)/c,\mathbfit{r}')}{u^{l+1}(r,z)}\biggr],
\end{eqnarray} 
where $J_1$ is the Bessel function of the first kind, $S(t, r)$ is the source function from (\ref{s}), $z=m_\varphi\sqrt{c^2(t-t')^2-|\mathbfit{r}-\mathbfit{r}'|^2}$, and $u(r, z)=\sqrt{1+(z/m_\varphi r)^2}$. Here the integration region $N$
is taken over the near zone, and $|\mathbfit{r}'|\ll|\mathbfit{r}|$. Substituting the source term $S(t, r)$ in the explicit form (\ref{s}) to (\ref{fi}) and performing the integration over $\mathbfit{r}'$ we find $\varphi$ in terms of the scalar multipole moments $\mathcal{M}^L_l$:
\begin{equation}\label{fi1} 
\varphi=\frac{4c_\varphi}{rc^2}\int_0^\infty dz J_1(z)\sum_{l=0}^\infty\frac{1}{c^ll!}n_L\partial^l_t\mathcal{M}^L_l,
\end{equation} 
where
\begin{eqnarray}\label{mll} 
\mathcal{M}^L_l&=&\mathcal{M}^{i_1i_2...i_l}_l(t, r, z)=\sum_a\biggl(M_a(t-r/c)r^L_a(t-r/c)\nonumber\\
&&-u^{-(l+1)}(r,z)M_a(t-ru(r,z)/c)r^L_a(t-ru(r,z)/c)\biggr)\nonumber\\
\end{eqnarray} 
and
\begin{eqnarray}\label{m} 
M_a(t)=&&m_a\biggl[1-2\frac{G_{4(0,0)}}{G_{4(1,0)}}\frac{s_a}{\phi_0}-\frac{v_a^2}{2c^2}\biggl(1-\frac{2G_{4(0,0)}}{G_{4(1,0)}}\frac{s_a}{\phi_0}\biggr)\nonumber\\
&-&3\sum_{b\neq a}\frac{m_b}{r_{ab}(t)c^2G_{4(0,0)}}\biggl(1-\frac{2G_{4(0,0)}}{G_{4(1,0)}}\frac{s_b}{\phi_0}\biggr)\nonumber\\
&+&\frac{6G_{4(1, 0)}c_\varphi}{c^2 G_{4(0, 0)}}\sum_{b\neq a}\frac{m_b}{r_{ab}(t)}e^{-m_\varphi R}\biggl(1-\frac{2G_{4(0,0)}}{G_{4(1,0)}}\frac{s_b}{\phi_0}\biggr)\nonumber\\
&-&\sum_{b\neq a}\frac{m_b}{r_{ab}(t)c^2}e^{-m_\varphi R}\biggl(1-\frac{2G_{4(0,0)}}{G_{4(1,0)}}\frac{s_b}{\phi_0}\biggr)\nonumber\\
&\times&\biggl(\frac{8c_\varphi s_a}{\phi_0}-\frac{8}{\phi_0}\frac{2G_{4(0,0)}}{G_{4(1,0)}}(s_a'-s_a^2+s_a)c_\varphi\biggr)\biggr].
\end{eqnarray} 
Here $n_L=n_{i_1}n_{i_2}...n_{i_l}$, $r^L_a(t)=r_a^{i_1}(t)r_a^{i_2}(t)...r_a^{i_l}(t)$.

The average rate of the binding energy change due to the scalar radiation is given by
\begin{equation}\label{scalar flux} 
\left<\dot E_\varphi\right> =-cr^2\int d\Omega \left<t^{0r}(\varphi)\right>=\frac{c^5r^2}{16\pi}\int d\Omega \ \left<\frac{G_{4(1,0)}}{2G_{4(0,0)}c_\varphi}\ \partial_0 \varphi\partial_r\varphi\right>.
\end{equation} 
In the case of the scalar energy flux we cannot change derivatives from spatial to temporal as in the tensor case because $\varphi$ is not a function of the argument $(t-r/c)$. Dependence upon $r$ is more complicated. This occurs due to the presence of the scalar field mass in equation~(\ref{phi_1}). So in the explicit form the temporal and spatial derivatives of $\varphi$ are
\begin{equation}\label{derivatives phi1} 
\partial_0\varphi=\frac{4c_\varphi}{rc^2}\int_0^\infty dz J_1(z)\sum_{l=0}^\infty\frac{1}{c^{l+1}l!}n_L\partial^{l+1}_t\mathcal{M}^L_l,
\end{equation} 
\begin{equation}\label{derivatives phi} 
\partial_r\varphi=-\frac{4c_\varphi}{rc^2}\int_0^\infty dz J_1(z)\sum_{l=0}^\infty\frac{1}{c^{l+1}l!}n_L\partial^{l+1}_t\mathcal{M}^L_{l+1},
\end{equation} 
where a scalar multipole moments $\mathcal{M}^L_{l+1}$ are defined as
\begin{eqnarray}\label{mll+1} 
\mathcal{M}^L_{l+1}&=&\mathcal{M}^{i_1i_2...i_l}_{l+1}(t, r, z)=\sum_a\biggl(M_a(t-r/c)r^L_a(t-r/c)\\
&&-u^{-(l+2)}(r,z) M_a(t-ru(r,z)/c)r^L_a(t-ru(r,z)/c)\biggr).\nonumber
\end{eqnarray} 

There are monopole, dipole, and dipole-octupole radiations in the scalar sector, besides quadrupole one, and the average rate of the binding energy change due to scalar radiation in terms of scalar multipole moments takes the form
\begin{eqnarray}\label{energy moments} 
\left<\dot E_\varphi\right> =&-& \frac{2c^5G_{4(1,0)}c_\varphi}{G_{4(0,0)}}\int dz_1dz_2J_1(z_1)J_2(z_2) \biggl\langle \frac{1}{c^6}\mathcal{\dot M}_0\mathcal{\dot M}_1\nonumber\\
&+&\frac{1}{6c^8}\biggl(2\mathcal{\ddot M}_1^k\mathcal{\ddot M}_2^k+\mathcal{\dot M}_0\mathcal{\dddot M}_3^{kk}+\mathcal{\dot M}_1\mathcal{\dddot M}_2^{kk}\biggr)\nonumber\\
&+&\frac{1}{60c^{10}}\biggl(2\mathcal{\dddot M}_2^{kl}\mathcal{\dddot M}_3^{kl}+\mathcal{\dddot M}_2^{kk}\mathcal{\dddot M}_3^{ll}\biggr)\nonumber\\
&+&\frac{1}{30c^{10}}\biggl(\mathcal{\ddot M}_1^{k}\mathcal{\ddddot M}_4^{kll}+\mathcal{\ddot M}_2^{k}\mathcal{\ddddot M}_3^{kll}\biggr)\biggr\rangle,
\end{eqnarray} 
where we have used the identity 
\begin{eqnarray}\label{energy moments} 
&\int \frac{d\Omega}{4\pi}n_{i_1}n_{i_2}...n_{i_k}=\begin{cases}
0, & \text{for $k=$odd,} \\
\frac{\delta_{i_1i_2}\delta_{i_3i_4}...\delta_{i_{k-1}i_k}+...}{(k+1)!!}, & \text{for $k=$even,}\nonumber
\end{cases}
\end{eqnarray} 
the final dots denote all possible pairing of indices.

Now we obtain the time derivatives of monopole, dipole, quadrupole, and octupole scalar moments for the quasi-circular orbit (\ref{quasicircular}):

1. Monopole.
\begin{equation}\label{monopole} 
\mathcal{\dot M}_0=\mathcal{\dot M}_1=0.
\end{equation} 

2. Dipole.
\begin{eqnarray}\label{dipole} 
\mathcal{\ddot M}_1^k&=&\mu R\omega^2\biggl(A_d+\bar{A}_{d}\frac{\mu}{c^2R}\biggr)\nonumber\\
&&\times[\cos(\omega(t-r/c))-u^{-2}\cos(\omega(t-ru/c)),\nonumber \\
&&\ \sin(\omega(t-r/c))-u^{-2}\sin(\omega(t-ru/c)), 0],\nonumber\\
\mathcal{\ddot M}_2^k&=&\mu R\omega^2\biggl(A_d+\bar{A}_{d}\frac{\mu}{c^2R}\biggr)\nonumber\\
&&\times[\cos(\omega(t-r/c))-u^{-3}\cos(\omega(t-ru/c)), \nonumber\\
&&\ \sin(\omega(t-r/c))-u^{-3}\sin(\omega(t-ru/c)), 0].\nonumber\\
\end{eqnarray} 

3. Quadrupole.
\begin{eqnarray}\label{quadrupole} 
\mathcal{\dddot M}_2^{kl}=\begin{pmatrix} \mathcal{\dddot M}_2^{11} & \mathcal{\dddot M}_2^{12} & 0 \\ \mathcal{\dddot M}_2^{12} & -\mathcal{\dddot M}_2^{11} & 0\\
0 & 0 & 0 \end{pmatrix},
\mathcal{\dddot M}_3^{kl}=\begin{pmatrix} \mathcal{\dddot M}_3^{11} & \mathcal{\dddot M}_3^{12} & 0 \\ \mathcal{\dddot M}_3^{12} & -\mathcal{\dddot M}_3^{11} & 0\\
0 & 0 & 0 \end{pmatrix},
\end{eqnarray} 
where 
\begin{eqnarray}\label{components} 
\mathcal{\dddot M}_2^{11}&=&4A_q\mu\omega^3R^2[\sin(2\omega(t-r/c))-u^{-3}\sin(2\omega(t-ru/c))],\nonumber\\
\mathcal{\dddot M}_2^{12}&=&-4A_q\mu\omega^3R^2[\cos(2\omega(t-r/c))-u^{-3}\cos(2\omega(t-ru/c))],\nonumber\\
\mathcal{\dddot M}_3^{11}&=&4A_q\mu\omega^3R^2[\sin(2\omega(t-r/c))-u^{-4}\sin(2\omega(t-ru/c))],\nonumber\\
\mathcal{\dddot M}_3^{12}&=&-4A_q\mu\omega^3R^2[\cos(2\omega(t-r/c))-u^{-4}\cos(2\omega(t-ru/c))]\nonumber.
\end{eqnarray} 

3. Octupole.
\begin{eqnarray}\label{octupole} 
\mathcal{\ddddot M}_3^{1kk}&=&A_o\mu\omega^4R^3[\cos(\omega(t-r/c))-u^{-4}\cos(\omega(t-ru/c))],\nonumber\\
\mathcal{\ddddot M}_3^{2kk}&=&A_o\mu\omega^4R^3[\sin(\omega(t-r/c))-u^{-4}\sin(\omega(t-ru/c))],\nonumber\\
\mathcal{\ddddot M}_4^{1kk}&=&A_o\mu\omega^4R^3[\cos(\omega(t-r/c))-u^{-5}\cos(\omega(t-ru/c))],\nonumber\\
\mathcal{\ddddot M}_4^{2kk}&=&A_o\mu\omega^4R^3[\sin(\omega(t-r/c))-u^{-5}\sin(\omega(t-ru/c))].\nonumber
\end{eqnarray} 
We use the following definitions:
\begin{eqnarray}\label{Ad}
A_d=&&\frac{2G_{4(0,0)}(s_2-s_1)}{G_{4(1,0)}\phi_0},\nonumber\\
A_q=&&1-\frac{2G_{4(0,0)}}{G_{4(1,0)}\phi_0}\frac{s_2m_1+s_1m_2}{m},\nonumber\\ 
A_o=&&\frac{m_1-m_2}{m}-\frac{2G_{4(0,0)}}{G_{4(1,0)}\phi_0}\frac{s_2m_1^2-s_1m_2^2}{m^2}
\end{eqnarray} 
and
\begin{eqnarray}\label{adbar}
\bar{A}_{d}=&-&\frac{7}{2G_{4(0,0)}}\biggl(\frac{m_2}{m_1}-\frac{m_1}{m_2}\biggr)+\frac{7}{G_{4(1,0)}\phi_0}\biggl(\frac{m_2s_1}{m_1}-\frac{m_1s_2}{m_2}\biggr)\nonumber\\
&+&\frac{6}{G_{4(1,0)}\phi_0}(s_1-s_2)+\frac{23}{4}c_\varphi\frac{G_{4(1,0)}}{G_{4(0,0)}}\biggl(\frac{m_2}{m_1}-\frac{m_1}{m_2}\biggr)\nonumber\\
&+&\frac{15c_\varphi}{2\phi_0}\biggl(\frac{m_1s_2}{m_2}-\frac{m_2s_1}{m_1}\biggr)+\frac{12c_\varphi}{\phi_0}\biggl(\frac{m_1s_1}{m_2}-\frac{m_2s_2}{m_1}\biggr)\nonumber\\
&+&\frac{14G_{4(0,0)}s_1s_2c_\varphi}{G_{4(1,0)}\phi_0^2}\biggl(\frac{m_2}{m_1}-\frac{m_1}{m_2}\biggr)+\frac{c_\varphi(s_1+s_2)}{2\phi_0}\nonumber\\
&\times&\biggl(\frac{m_2}{m_1}-\frac{m_1}{m_2}\biggr)+\frac{8G_{4(0,0)}c_\varphi}{G_{4(1,0)}\phi_0^2}\biggl(\frac{m_2s_1}{m_1}-\frac{m_1s_2}{m_2}\biggr)\nonumber\\
&+&\frac{8G_{4(0,0)}c_\varphi(s_1-s_2)}{G_{4(1,0)}\phi_0^2}+\frac{9G_{4(0,0)}c_\varphi}{G_{4(1,0)}\phi_0^2}\biggl(\frac{s^2_2m_1}{m_2}-\frac{s^2_1m_2}{m_1}\biggr)\nonumber\\
&+&\frac{4c_\varphi(s_1-s_2)}{\phi_0}+\frac{18G^2_{4(0,0)}c_\varphi}{G^2_{4(1,0)}\phi_0^3}\biggl(\frac{s^2_1s_2m_2}{m_1}-\frac{s^2_2s_1m_1}{m_2}\biggr)\nonumber\\
&+&\frac{16G^2_{4(0,0)}c_\varphi}{G^2_{4(1,0)}\phi_0^3}(s^2_1s_2-s^2_2s_1)-\frac{16G^2_{4(0,0)}c_\varphi}{G^2_{4(1,0)}\phi_0^3}(s_1s_2'-s_2s_1')\nonumber\\
&-&\frac{8G_{4(0,0)}c_\varphi}{G_{4(1,0)}\phi_0^2}\biggl(\frac{m_2s_1'}{m_1}-\frac{m_1s_2'}{m_2}\biggr)+\frac{8G_{4(0,0)}c_\varphi}{G_{4(1,0)}\phi_0^2}(s_2^2-s_1^2)\nonumber\\
&-&\frac{8G_{4(0,0)}c_\varphi}{G_{4(1,0)}\phi_0^2}(s_1'-s_2')+\frac{16G^2_{4(0,0)}c_\varphi s_1s_2}{G^2_{4(1,0)}\phi_0^3}\biggl(\frac{m_1}{m_2}-\frac{m_2}{m_1}\biggr)\nonumber\\
&-&\frac{16G^2_{4(0,0)}c_\varphi}{G^2_{4(1,0)}\phi_0^3}\biggl(\frac{s_1s_2'm_1}{m_2}-\frac{s_2s_1'm_2}{m_1}\biggr).
\end{eqnarray}

	We can divide the scalar sector on the dipole, quadrupole and dipole-octupole components:
	\begin{equation}\label{components} 
	\left<\dot E_\varphi\right>=\left<\dot E^D_\varphi\right>+\left<\dot E^Q_\varphi\right>+\left<\dot E^{DO}_\varphi\right>,
	\end{equation} 
	where the scalar dipole part is
	\begin{eqnarray}\label{ed} 
	\left<\dot E^D_\varphi\right>=& - &\frac{2G_{4(1,0)}c_\varphi}{3c^3G_{4(0,0)}}\int\int dz_1dz_2J_1(z_1)J_2(z_2)\nonumber \\
	&\times&\biggl\langle \mathcal{\ddot M}_1^k(z_1)\mathcal{\ddot M}_2^k(z_2)\biggr\rangle=- \frac{2G_{4(1,0)}c_\varphi}{3c^3G_{4(0,0)}}\frac{\mu^2\mathcal{G}^2_{12}m^2 }{R^4}\nonumber\\
	&\times&\biggl(A_d^2+A_d\bar{A}_d\frac{2\mu}{c^2R}\biggr)\bigl[1-\cos(\omega r/c)\langle\cos(\omega ru/c)\rangle_2\nonumber\\
	&-&\sin(\omega r/c)\langle\sin(\omega ru/c)\rangle_2-(\cos(\omega r/c)\nonumber\\
	&-&\langle\cos(\omega ru/c)\rangle_2)\langle\cos(\omega ru/c)\rangle_3-(\sin(\omega r/c)\nonumber\\
	&-&\langle\sin(\omega ru/c)\rangle_2)\langle\sin(\omega ru/c)\rangle_3\bigr],
	\end{eqnarray} 
	the scalar quadrupole part is
	\begin{eqnarray}\label{eq} 
	\left<\dot E^Q_\varphi\right>=& - &\frac{G_{4(1,0)}c_\varphi}{15c^5G_{4(0,0)}}\int\int dz_1dz_2J_1(z_1)J_2(z_2)\biggl\langle\mathcal{\dddot M}_2^{kl}\mathcal{\dddot M}_3^{kl}\biggr\rangle\nonumber\\
	=&-& \frac{32G_{4(1,0)}c_\varphi}{15c^5G_{4(0,0)}}\frac{\mu^2\mathcal{G}^3_{12}m^3 }{R^5}A_q^2\nonumber\\
	&\times&\bigl[1-\cos(2\omega r/c)\langle\cos(2\omega ru/c)\rangle_3\nonumber\\
	&-&\sin(2\omega r/c)\langle\sin(2\omega ru/c)\rangle_3-(\cos(2\omega r/c)\nonumber\\
	&-&\langle\cos(2\omega ru/c)\rangle_3)\langle\cos(2\omega ru/c)\rangle_4-(\sin(2\omega r/c)\nonumber\\
	&-&\langle\sin(2\omega ru/c)\rangle_3)\langle\sin(2\omega ru/c)\rangle_4\bigr],
	\end{eqnarray} 
	and the scalar dipole-octupole part is
	\begin{eqnarray}\label{edo} 
	\left<\dot E^{DO}_\varphi\right>=& - &\frac{G_{4(1,0)}c_\varphi}{15c^5G_{4(0,0)}}\int\int dz_1dz_2J_1(z_1)J_2(z_2) \nonumber\\
	&\times&\biggl\langle\biggl(\mathcal{\ddot M}_1^{k}\mathcal{\ddddot M}_4^{kll}+\mathcal{\ddot M}_2^{k}\mathcal{\ddddot M}_3^{kll}\biggr)\biggr\rangle\nonumber\\
	=&&\frac{G_{4(1,0)}c_\varphi}{15c^5G_{4(0,0)}}\frac{\mu^2\mathcal{G}^3_{12}m^3 }{R^5}A_dA_o\nonumber\\
	&\times&\bigl[2-\cos(\omega r/c)\bigl(\langle\cos(\omega ru/c)\rangle_2+\langle\cos(\omega ru/c)\rangle_3\nonumber\\
	&+&\langle\cos(\omega ru/c)\rangle_4+\langle\cos(\omega ru/c)\rangle_5\bigr)-\sin(\omega r/c)\nonumber\\
	&\times&\bigl(\langle\sin(\omega ru/c)\rangle_2+\langle\sin(\omega ru/c)\rangle_3+\langle\sin(\omega ru/c)\rangle_4\nonumber\\
	&+&\langle\sin(\omega ru/c)\rangle_5\bigr)+\langle\cos(\omega ru/c)\rangle_2)\langle\cos(\omega ru/c)\rangle_5\nonumber\\
	&+&\langle\cos(\omega ru/c)\rangle_3)\langle\cos(\omega ru/c)\rangle_4\nonumber\\
	&+&\langle\sin(\omega ru/c)\rangle_2)\langle\sin(\omega ru/c)\rangle_5\nonumber\\
	&+&\langle\sin(\omega ru/c)\rangle_3)\langle\sin(\omega ru/c)\rangle_4\bigr].
	\end{eqnarray} 
	Here we have used the Kepler's third law (\ref{Kepler}) and introduced the following notation for integrals:
	\begin{equation}\label{sin_integrals} 
	\left<\sin\biggl(\frac{\omega ru}{c}\biggr)\right>_n\equiv\int^{\infty}_0\sin\biggl(\frac{\omega r}{c}\sqrt{1+\biggl(\frac{z}{m_\varphi r}\biggr)^2}\biggr)\frac{J_1(z)dz}{\biggl[1+\biggl(\frac{z}{m_\varphi r}\biggr)^2\biggr]^{n/2}},
	\end{equation} 
	\begin{equation}\label{cos_integrals} 
	\left<\cos\biggl(\frac{\omega ru}{c}\biggr)\right>_n\equiv \int^{\infty}_0\cos\biggl(\frac{\omega r}{c}\sqrt{1+\biggl(\frac{z}{m_\varphi r}\biggr)^2}\biggr)\frac{J_1(z)dz}{\biggl[1+\biggl(\frac{z}{m_\varphi r}\biggr)^2\biggr]^{n/2}}.
	\end{equation} 
Let us obtain the total power of scalar radiation and perform these integrals in the limit $r\to\infty$. The detailed calculations can be found in the papers of \citet{massivebd, zhang}. We  present only the final result as follows:
\begin{align}\label{integrals} 
\lim_{r\to\infty}\left<\sin\biggl(\frac{\omega ru}{c}\biggr)\right>_n\equiv\begin{cases}
\sin(\frac{\omega r}{c})-\biggl(\frac{v_{\varphi}(\omega)}{c}\biggr)^{n-1}\cos(\omega rv_{\varphi}(\omega)), \\
 \text{for $\omega>cm_\varphi$,} \\
\\
\sin(\frac{\omega r}{c})-\frac{(-1)^{n-1}-1}{2}\biggl(\frac{v_{\varphi}(\omega)}{c}\biggr)^{n-1}e^{-i\omega rv_{\varphi}(\omega)}, \\
 \text{for $\omega<cm_\varphi$,}
\end{cases}
\end{align} 
and
\begin{align}\label{integrals_2} 
\lim_{r\to\infty}\left<\cos\biggl(\frac{\omega ru}{c}\biggr)\right>_n\equiv\begin{cases}
\cos(\frac{\omega r}{c})-\biggl(\frac{v_{\varphi}(\omega)}{c}\biggr)^{n-1}\cos(\omega rv_{\varphi}(\omega)),\\
\text{for $\omega>cm_\varphi$,} \\
\\
\cos(\frac{\omega r}{c})-\frac{(-1)^{n-1}+1}{2}\biggl(\frac{v_{\varphi}(\omega)}{c}\biggr)^{n-1}e^{-i\omega rv_{\varphi}(\omega)},\\
\text{for $\omega<cm_\varphi$,}
\end{cases}
\end{align} 
where $v_{\varphi}(\omega)=c\sqrt{1-m_\varphi^2c^2/\omega^2}$ is the propagation speed of the scalar gravitational radiation.

Finally, we obtain the expression for the dipole
\begin{eqnarray}\label{dipole part} 
\left<\dot E^D_\varphi\right>=&-&\frac{2G_{4(1,0)}c_\varphi}{3c^3G_{4(0,0)}}\frac{\mu^2\mathcal{G}^2_{12}m^2 }{R^4}\biggl(A_d^2+A_d\bar{A}_d\frac{2\mu}{c^2R}\biggr) \biggl(\frac{v_{\varphi}(\omega)}{c}\biggr)^3\nonumber\\
&\times&\Theta(\omega-cm_\varphi),
\end{eqnarray} 
the quadrupole
\begin{equation}\label{quadrupole part} 
\left<\dot E^Q_\varphi\right>=-\frac{32G_{4(1,0)}c_\varphi}{15c^5G_{4(0,0)}}\frac{\mu^2\mathcal{G}^3_{12}m^3 }{R^5}A_q^2 \biggl(\frac{v_{\varphi}(2\omega)}{c}\biggr)^5\Theta(2\omega-cm_\varphi),
\end{equation} 
and the dipole-octupole 
\begin{equation}\label{quadrupole part} 
\left<\dot E^{DO}_\varphi\right>=\frac{G_{4(1,0)}c_\varphi}{15c^5G_{4(0,0)}}\frac{\mu^2\mathcal{G}^3_{12}m^3 }{R^5}A_dA_o \biggl(\frac{v_{\varphi}(\omega)}{c}\biggr)^5\Theta(\omega-cm_\varphi)
\end{equation} 
parts of the scalar radiation power. Here $\Theta(\omega-cm_\varphi)$ is the Heaviside function.

It is important to emphasize that in our work we take into account PN corrections to the dipole term and dipole-octupole term which are absent if we neglect two terms of order $O(1/c^4)$ during considering the expression for $\varphi$. Thus, our consideration of the scalar radiation power is complete. 

The total power of the scalar radiation is 
\begin{eqnarray}\label{total} 
\left<\dot E_\varphi\right>=&&\left<\dot E^Q_g\right>+\left<\dot E^D_\varphi\right>+\left<\dot E^Q_\varphi\right>+\left<\dot E^{DO}_\varphi\right>\nonumber\\
=&-&\frac{32\mu^2(\mathcal{G}_{12}m)^3}{5c^5G_{4(0,0)}R^5}\biggl[1+ \frac{5c^2G_{4(1,0)}c_\varphi R}{48\mathcal{G}_{12}m}\nonumber\\
&\times&\biggl(A_d^2+A_d\bar{A}_{d}\frac{2\mu}{c^2R}\biggr) \biggl(\frac{v_{\varphi}(\omega)}{c}\biggr)^3\Theta(\omega-cm_\varphi)\nonumber\\
&+&\frac{G_{4(1,0)}c_\varphi}{3}A_q^2 \biggl(\frac{v_{\varphi}(2\omega)}{c}\biggr)^5\Theta(2\omega-cm_\varphi)\nonumber\\
&-&\frac{G_{4(1,0)}c_\varphi}{96}A_dA_o\biggl(\frac{v_{\varphi}(\omega)}{c}\biggr)^5\Theta(\omega-cm_\varphi)\biggr].
\end{eqnarray} 
According to the equation~(\ref{total})  in contrast to all other contributions, the dipole-octupole term has the opposite sign and describes the negative modification of the energy flux at the same PN order as the quadrupole radiation contribution.

Using equations~(\ref{energy}) and (\ref{energy_losses}) we  find the final form of the orbital decay rate including the tensor and scalar parts:
\begin{eqnarray}\label{orb_dec} 
\frac{\dot P_b^{th}}{P_b}=&-&\frac{96\mu(\mathcal{G}_{12}m)^2}{5c^5G_{4(0,0)}R^4}\biggl[1+ \frac{5c^2G_{4(1,0)}c_\varphi R}{48\mathcal{G}_{12}m}\nonumber\\
&\times&\biggl(A_d^2+A_d\bar{A}_{d}\frac{2\mu}{c^2R}\biggr) \biggl(\frac{v_{\varphi}(\omega)}{c}\biggr)^3\Theta(\omega-cm_\varphi)\nonumber\\
&+&\frac{G_{4(1,0)}c_\varphi}{3}A_q^2 \biggl(\frac{v_{\varphi}(2\omega)}{c}\biggr)^5\Theta(2\omega-cm_\varphi)\nonumber\\
&-&\frac{G_{4(1,0)}c_\varphi}{96}A_dA_o\biggl(\frac{v_{\varphi}(\omega)}{c}\biggr)^5\Theta(\omega-cm_\varphi)\biggr],
\end{eqnarray} 
where index "th" denotes expression obtained in the frameworks of considering subclass of the Horndeski gravity. According to the equation~(\ref{orb_dec}) the main contribution to scalar radiation is produced by scalar dipole term. Also from expressions~(\ref{Ad}) and (\ref{adbar}) one can see that the contribution of the scalar dipole radiation depends upon the difference ($s_1-s_2$). Thus, the scalar dipole radiation is the most noticeably in mixed binaries where this difference in sensitivities reaches the maximum values.

Here and further we take into account that $m_\varphi<\omega/c$, thus $\Theta(\omega-cm_\varphi)=1$. Using the Kepler's third law (\ref{Kepler}) (in the quasi-circular orbit $a=R$) and explicit form of the scalar propagation speed $v_{\varphi}(\omega)$, equation~(\ref{orb_dec}) can be rewritten as
\begin{eqnarray}\label{first} 
\frac{\dot P_b^{th}}{\dot P_b^{GR}}=&&\frac{\mathcal{G}_{12}^{\frac{2}{3}}}{G^{\frac{5}{3}}G_{4(0,0)}}\biggl\{1+ \frac{5G_{4(1,0)}c_\varphi}{48}\biggl(\frac{P_b c^3}{2\pi m \mathcal{G}_{12}}\biggr)^{\frac{2}{3}}\nonumber\\ 
&\times&\biggl[A_d^2+\frac{2\mu}{c^2} A_d\bar{A}_{d}\biggl(\frac{4\pi^2}{P_b^2 m \mathcal{G}_{12}}\biggr)^{\frac{1}{3}}\biggr]\biggl(1-\frac{m_\varphi^2c^2P_b^2}{4\pi^2}\biggr)^{\frac{3}{2}}\nonumber\\ 
&+& \frac{G_{4(1,0)}c_\varphi}{3}A_q^2 \biggl(1-\frac{m_\varphi^2c^2P_b^2}{16\pi^2}\biggr)^{\frac{5}{2}}\nonumber\\ 
&-&\frac{G_{4(1,0)}c_\varphi}{96}A_dA_o \biggl(1-\frac{m_\varphi^2c^2P_b^2}{4\pi^2}\biggr)^{\frac{5}{2}}\biggr\},
\end{eqnarray} 
where $\dot P_b^{GR}$ is the value of orbital decay predicted by GR:
\begin{equation}\label{GR} 
\dot P_b^{GR}=-\frac{192\pi\mu}{5c^5m}\biggl(\frac{2\pi Gm}{P_b}\biggr)^{\frac{5}{3}}.
\end{equation} 
Readers who wish to familiarize with the full expression of the orbital decay within the framework of the standard massive scalar-tensor theory after all substitutions can find it in \hyperref[Appendix ]{Appendix A}.

\section{Observational constraints}\label{sec:Observational constrains on Horndeski gravity from binary pulsars}

At the moment, the GR perfectly describes all observational data from binary pulsars within observational uncertainties \citep{tw, stairs, kramer1, bhat, 2012a,   17381,  ransom, pulsars, archi}. Therefore, all deviations from GR predicted by modified gravity should be smaller than existing observational uncertainties. This fact allows to obtain very strict constraints on the considering subclass of the Horndeski gravity.

The observational value of the orbital period change $\dot P_b^{obs}$ consists of various components which have the different nature: intrinsic and kinematic effects \citep{damourtaylor, kinematic}. We are interested in the intrinsic part $\dot P_b^{intr}$ because the dominant element of this component is the orbital period change due to the emission of gravitational waves. Also, intrinsic part includes effects of the mass loss from the binary and from tidal torques \citep{damourtaylor, kinematic} but at the current stage we consider only such systems where these effects are negligibly small in relation to the effect of gravitational radiation.

The constraints on the Horndeski gravity can be obtained from the comparing of the predicted quantity $\dot P_b^{th}/\dot P_b^{GR}$ and the observational quantity $\dot P_b^{intr}/\dot P_b^{GR}$ at 95\% confidence level:
\begin{equation}\label{deviation} 
\biggl|\cfrac{\dot P_b^{th}}{\dot P_b^{GR}}-\cfrac{\dot P_b^{intr}}{\dot P_b^{GR}}\biggr|\leq2\sigma,
\end{equation} 
where $\sigma$ is the observational uncertainty.

\subsection{Constraints on the massive scalar-tensor theories}

The scalar dipole radiation prevails in the predictions of the standard massive scalar-tensor theories for the orbital period decay. The contribution of the scalar dipole part is the most noticeable in the mixed binary systems \citep{will1, will11, sens, will12}. We test the massive scalar-tensor models in mixed binary system PSR J1738+0333. This system has the most accurate observational data among quasi-circular mixed binaries. The mass of the white dwarf and mass ratio were obtained in theory-independent way (under the assumption that nonperturbative strong-field effects are absent and higher-order contributions in powers of the gravitational binding energies of the bodies can be neglected) \citep{1738, 17381}. This fact allows to test a theory of gravity using only one PPK parameter ($\dot P_b$). The orbital parameters for this system are listed in Table~\ref{tab:J1738+0333}.

\begin{table}
	\begin{minipage}[h]{.45\textwidth}
		\caption{Parameters PSR J1738+0333 \citep{1738, 17381}}\label{tab:J1738+0333}
		\begin{tabular}{lll} 
			
			\hline
			Parameter & Physical meaning & Experimental value\\
			\hline
			$P_b $ & orbital period & $ 0.3547907398724(13)$ day\\

			$e$ & eccentricity & $ 0.34(11) \times 10^{-6} $\\

			$\dot P_b^{obs}$ & observational secular & $ -0.170(31)\times 10^{-13}$\\
			& change of $P_b$ &\\
			
			$\dot P_b^{intr}$ & intrinsic secular & $ -0.259(32)\times 10^{-13}$\\
			& change of $P_b$ &\\
			
			$\dot P_b^{intr}/\dot P_b^{GR}$ & relation between $\dot P_b^{intr}$& $ 0.93(13) $\\
			& and $\dot P_b^{GR}$ &\\
			
			$m_1 $ & mass of the pulsar & $1.46^{+0.06}_{-0.05}\  M_{\bigodot}$\\
			
			$m_2 $ & mass of the white dwarf & $0.181^{+0.007}_{-0.005}\  M_{\bigodot}$\\
			
			$m $ & total system mass & $1.65^{+0.07}_{-0.06}\  M_{\bigodot}
			$\\
			
			\hline
		\end{tabular}
	\end{minipage}
	\end {table}
	
	The dipole radiation contributes as the leading order in the scalar sector of orbital period decay, so we can neglect other terms of scalar sector in the definition of the first derivative of the orbital period. Thus the orbital period change (\ref{first derivative of the orbital period}) takes the form:
	\begin{eqnarray}\label{first2} 
	\cfrac{\dot P_b^{th}}{\dot P_b^{GR}}=&&\cfrac{\mathcal{G}_{12}^{\frac{2}{3}}}{G^{\frac{5}{3}}G_{4(0,0)}}\biggl[1+ \cfrac{5c_\varphi}{12}\biggl(\cfrac{P_b c^3}{2\pi m \mathcal{G}_{12}}\biggr)^{\frac{2}{3}}\nonumber\\
&\times&\biggl(\cfrac{G^2_{4(0,0)}(s_{NS}-s_{WD})^2}{G_{4(1,0)}\phi^2_0}\biggr)\biggl(1-\cfrac{m_\varphi^2c^2P_b^2}{4\pi^2}\biggr)^{\frac{3}{2}}\biggr],\nonumber\\
	\end{eqnarray}
	where $s_{NS}$ is sensitivity of neutron star and $s_{WD}$ is the sensitivity of white dwarf.
	
	Using (\ref{deviation}) and (\ref{first2}) we obtain the following bounds on the considering subclass of the Horndeski theory:
	\begin{eqnarray}\label{wdns1} 
	&&\biggl|\cfrac{\mathcal{G}_{12}^{\frac{2}{3}}}{G^{\frac{5}{3}}G_{4(0,0)}}\biggl[1+ \cfrac{5c_\varphi}{12}\biggl(\cfrac{P_b c^3}{2\pi m \mathcal{G}_{12}}\biggr)^{\frac{2}{3}}\biggl(1-\cfrac{m_\varphi^2c^2P_b^2}{4\pi^2}\biggr)^{\frac{3}{2}}\nonumber\\ 
	&&\times\biggl(\cfrac{G^2_{4(0,0)}(s_{NS}-s_{WD})^2}{G_{4(1,0)}\phi^2_0}\biggr)\biggr]-0.93\biggr|\leq 0.26.
	\end{eqnarray}
	
	From the condition $m_{\varphi}<\omega/c$ one obtains the constraints on scalar field mass (see Table~\ref{tab:J1738+0333}):
	\begin{equation}\label{massnswd} 
	m_{\varphi}<7\times 10^{-15}(\text{cm}^{-1}).
	\end{equation}

	Generally speaking, the longer the system period, the more accurate constraints on the scalar field mass. However, for systems with  large orbital period the quantity $\dot P_b$ has very inaccurately measured value. Therefore, the best laboratories for testing scalar-tensor theories are systems combining both a large orbital period and a well-measured value of  the orbital period decay.
	
	In the general case of the scalar-tensor theories, the mixed binary system gives the better constraints than neutron star-neutron star binary due to the value of the difference of sensitivities. However, not all scalar-tensor models include the sensitivities. 
	If there is no concept of sensitivity in the theory, then in the expression for the predicted orbital period change $\dot P_b^{th}$ only quadrupole terms remain, regardless of the type of the binary pulsar. In this case the best restrictions could be found from the binary pulsar with the most accurate value of the quantity $\dot P_b^{obs}/\dot P_b^{GR}$. The double binary pulsar PSR J0737-3039 is such system \citep{kramer, kramer1}.
		
	The system PSR J0737-3039 is the only known double binary pulsar. The extraordinary closeness of the system components, small orbital period and the fact that we see almost edge-on system allow to investigate the manifestation of relativistic effects with the highest available precision. This system consists of two pulsars and provides the most accurate data among all binary pulsars. The observational data for this system is listed in the Table~\ref{tab:J0737-3039} \citep{kramer, kramer1}.

	\begin{table}
		\begin{minipage}{.45\textwidth}
			\caption{Parameters PSR J0737-3039 \citep{kramer1}}\label{tab:J0737-3039}
			\begin{tabular}{l l l} 
				\hline
				
				Parameter & Physical meaning & Experimental value\\
				\hline
				$P_b $ & orbital period & $ 0.10225156248(5)$ day\\

				$e$ & eccentricity & $ 0.0877775(9) $\\

				$\dot P_b^{obs}$ & observational secular & $ -1.252(17) \times 10^{-12}$\\
				& change of $P_b$ &\\
				
				$\dot P_b^{obs}/\dot P_b^{GR}$ & relation between $\dot P_b^{obs}$& $ 1.003(14) $\\
				& and $\dot P_b^{GR}$ &\\
				
				$m_1 $ & mass of the first pulsar & $1.3381(7)\  M_{\bigodot}$\\
				
				$m_2 $ & mass of the second pulsar & $1.2489(7)\  M_{\bigodot}$\\
				
				$m $ & total system mass & $2.58708(16)\  M_{\bigodot}
				$\\
				\hline
				
			\end{tabular}
		\end{minipage}
	\end{table}

	In PSR J0737-3039 the kinematic contribution in the orbital period decay is negligibly small \citep{kramer, kramer1}. Thus, the quatities $\dot P_b^{obs}$ and $\dot P_b^{intr}$  almost the same (within observable accuracy). Using method~(\ref{deviation}) and the observational data from Table~\ref{tab:J0737-3039}, it is possible to obtain the following constraints for the scalar-tensor theories without sensitivities:
	\begin{equation}\label{ns}
	\biggl|1.003-\cfrac{\mathcal{G}_{12}^{\frac{2}{3}}}{G^{\frac{5}{3}}G_{4(0,0)}}\biggl[1+ \cfrac{G_{4(1,0)}c_\varphi}{3}\bigl(1-4\times10^{26}m^2_{\varphi}\bigr)^{\frac{5}{2}}\biggr]\biggr|\leq 0.028.
	\end{equation} 
	
	It is important to emphasize that these bounds do not include the masses of the components, since the absence of sensitivities nullifies all such terms. So our estimation is correct regardless of the method used to obtain the masses of companions in the system.

	So the observational data of PSR J0737-3039 impose the following restrictions on scalar field mass: 
	\begin{equation}\label{massns} 
	m_{\varphi}<5 \times 10^{-14} (\text{cm}^{-1}).
	\end{equation}
	These restrictions are connected with the orbital period of the binary system and do not depend on the specific choice of the scalar-tensor theory. 
	
	\subsection{Scalar-tensor specific cases: hybrid f(R)-gravity and massive Brans-Dicke theory}
	
	The action (\ref{act}) with the set of the gravitational Lagrangian densities (\ref{Lagrangians2}) is the generic one for the standard massive scalar-tensor theories with the second order field equations. Taking different sets of $G_i$ one can obtain limiting cases of this theory \citep{kobayashi}. In this work we consider two mathematically very similar  but physically very different  specific massive scalar-tensor cases: the hybrid metric-Palatini f(R)-gravity \citep{hybrid, hybrid1} and massive Brans-Dicke theory \citep{bd,massivebd}. 
	
	Our starting point is the hybrid metric-Palatini f(R)-gravity. This theory is geometrical but can be reduced to scalar-tensor model as any other f(R)-gravity \citep{tey}. The hybrid f(R)-theory was created as a mixture of metric and Palatini approaches to eliminate the disadvantages of both of them. This model allows to describe the accelerated expansion of the Universe and the galaxies rotation curves in a purely geometric way without introducing new particles. An additional interesting aspect of hybrid f(R)-gravity is the possibility to generate long-range forces without  conflict with local tests of gravity and without invoking any kind of screening mechanism (which would however require that at the present time the cosmological evolution reduces to GR)\citep{hybrid1}. This theory has already been well studied in cosmology and in different galaxies \citep{hybrid, hybrid1}. However, it is necessary to test any theory of gravity in the different field limits. So we test the hybrid f(R)-gravity in a strong field of binary pulsars.
	
	To reduce the general scalar-tensor action  (\ref{act}) to hybrid f(R)-gravity, it is necessary to choose the following set of parameters $G_i$:
	\begin{equation}\label{hybrid} 
	G_2=-\cfrac{3X}{G\phi}-V(\phi),\ G_3=0,\ G_4=\cfrac{1+\phi}{G},\ G_5=0
	\end{equation} 
	and 
	\begin{equation}\label{hybrid1} 
	G_{4(0,0)}=\cfrac{1+\phi_0}{G}, \ G_{4(1,0)}=\cfrac{1}{G},\ G_{3(1,0)}=0,\ G_{2(0,1)}=-\cfrac{3}{G\phi_0}.
	\end{equation}

	The hybrid metric-Palatini f(R)-gravity is pure geometrical theory and sensitivities $s_a$ do not appear in this model. Thus, the scalar quadrupole term is leading one in both types of pulsars system: in mixed binaries and in neutron star-neutron star binaries.
	
\begin{figure}
		\begin{minipage}[h]{.45\textwidth}
			\includegraphics[width=\columnwidth]{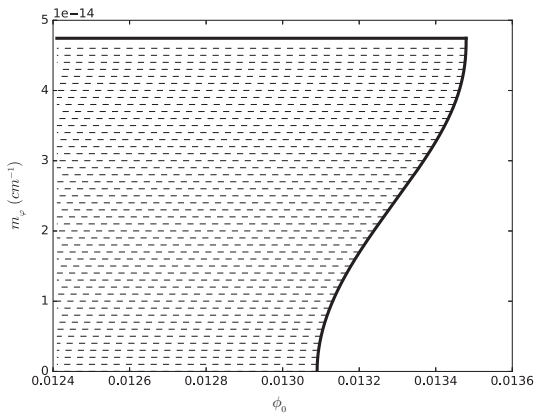}
			\caption{Hybrid f(R)-gravity. Dependence of the scalar field mass upon scalar field background value in the case of the system PSR J0737-3039. The dashed region corresponds to allowable values. The horizontal bold line is the critical value of scalar mass ($m_\varphi=2\omega/c$).}
			\label{fig:h0737}
		\end{minipage}
	\end{figure}

	The observational data from  PSR J0737-3039 provides the following constraints on the hydrid f(R)-gravity (from (\ref{ns})):
	\begin{equation}\label{hybrid2} 
	0.975\leq\cfrac{1}{(1+\phi_0)^{\frac{5}{3}}}\biggl(1-\cfrac{5\phi_0}{18}(1-2\times10^{26}m_\varphi^2)\biggr)\leq1.
	\end{equation} 
	The Fig.~\ref{fig:h0737} reflects the dependence $\phi_0$ from the scalar field mass $m_\varphi$ for the system PSR J0737-3039.

\begin{figure}
			\includegraphics[width=7cm]{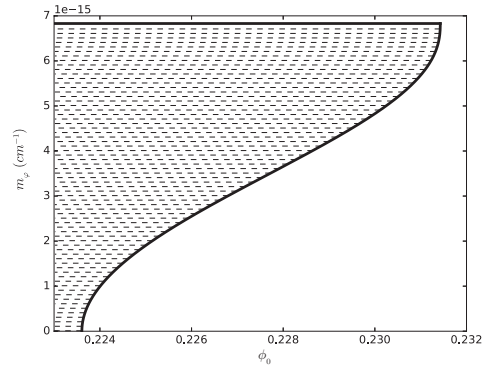}
			\caption{Hybrid f(R)-gravity. Dependence of the scalar field mass upon scalar field background value in the case of the system PSR J1738+0333. The dashed region corresponds to allowable values. The horizontal bold line is the critical value of scalar mass ($m_\varphi=2\omega/c$).}
			\label{fig:h1738}
	\end{figure}

	Since the theory does not contain sensitivity, the quarupole terms only contribute  to the value of orbital period decay. Thus, the mixed binary PSR J1738+0333 gives the following bounds (from the method~(\ref{deviation})) in the case of hybrid f(R)-gravity (see Fig.~\ref{fig:h1738}):
	\begin{equation}\label{hybrid3} 
	0.67\leq\cfrac{1}{(1+\phi_0)^{\frac{5}{3}}}\biggl(1-\cfrac{5\phi_0}{18}(1-3\times10^{27}m_\varphi^2)\biggr)\leq1.
	\end{equation} 
	
\begin{figure*}
		\begin{minipage}[h]{0.49\linewidth}
			{\includegraphics[width=8cm]{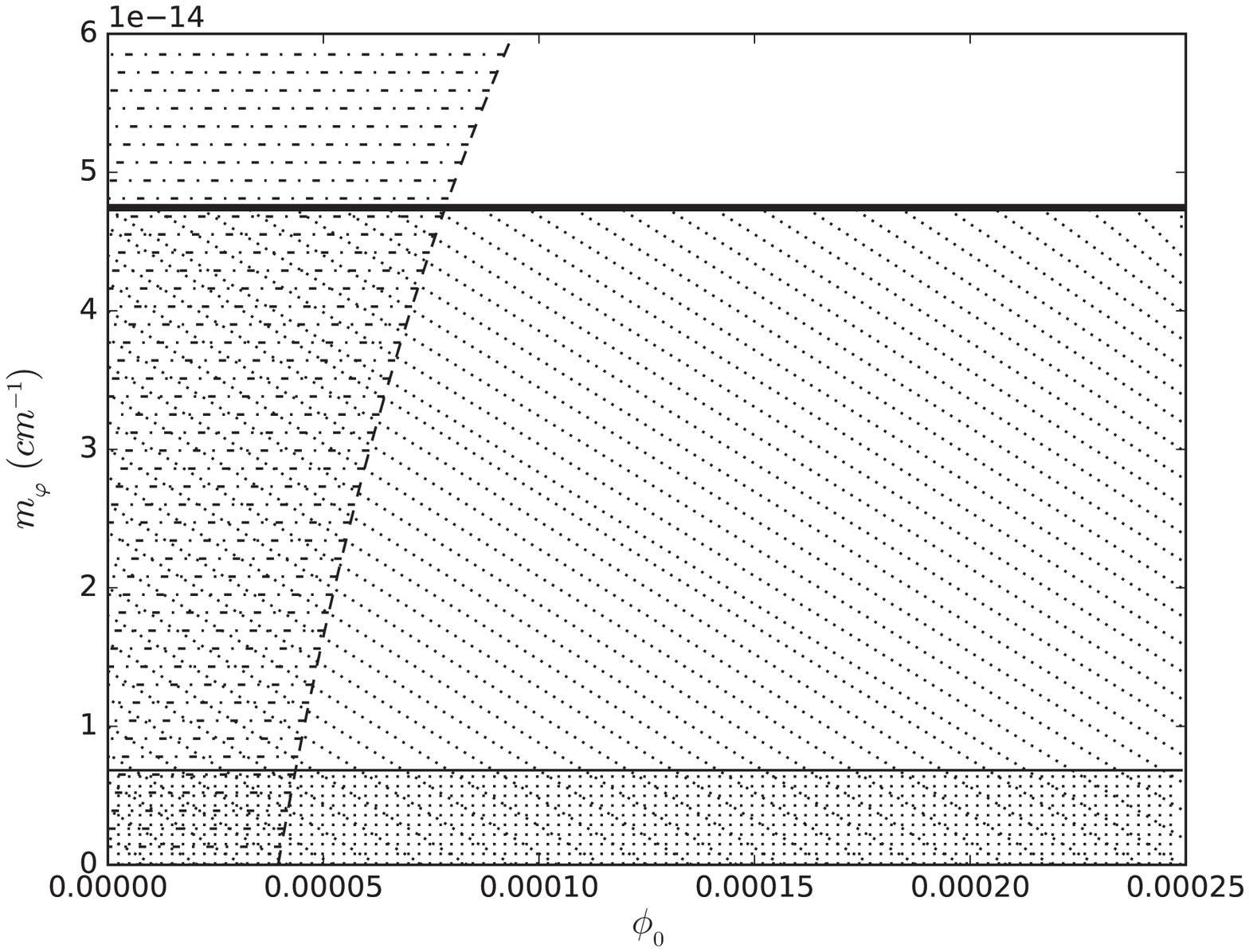}  \\ a)
			} 
		\end{minipage}
		\hfill
		\begin{minipage}[h]{0.49\linewidth}
			{\includegraphics[width=8cm]{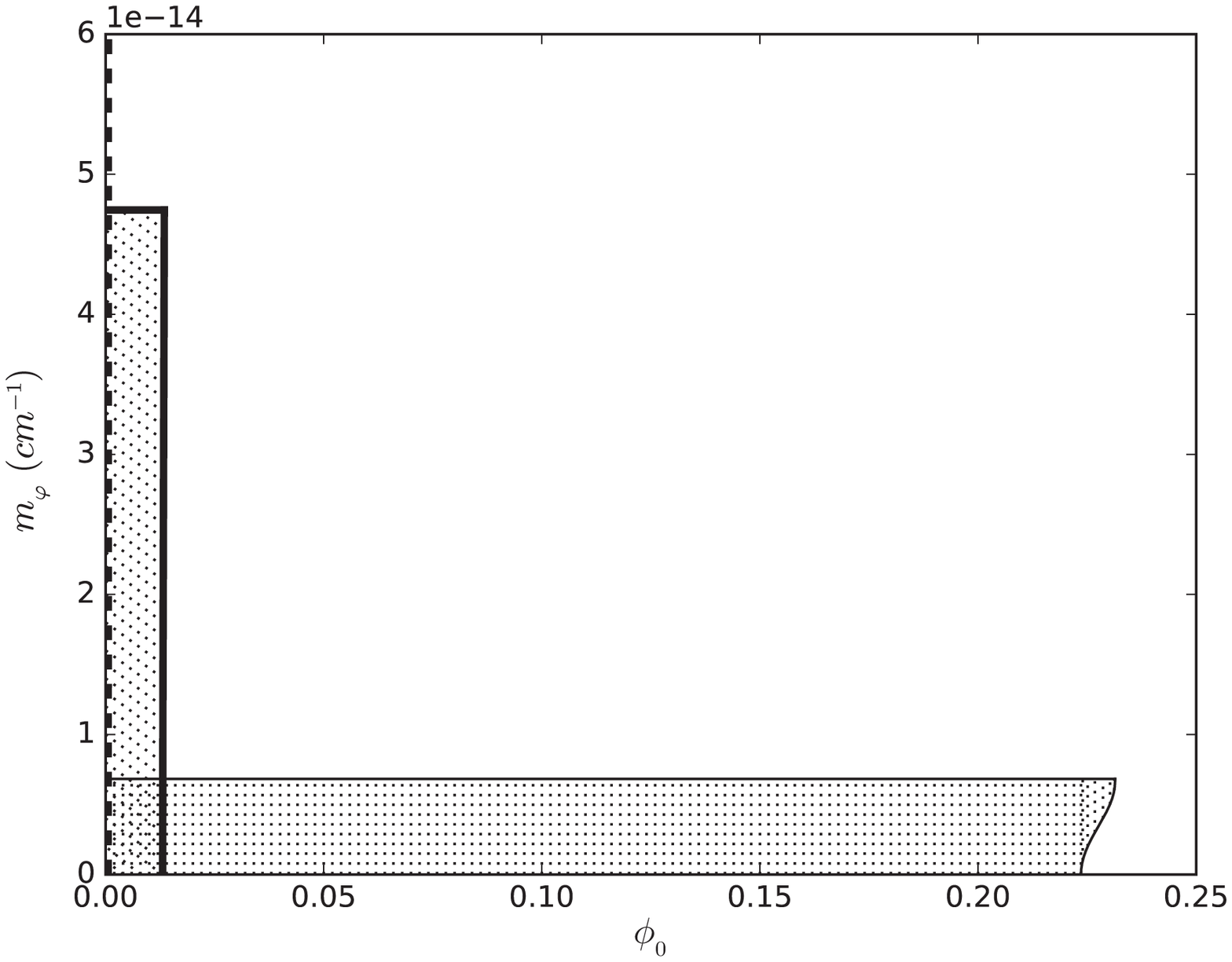} \\ b)
			}
		\end{minipage}
		\caption{Hybrid f(R)-gravity. Dependence of the scalar field mass upon scalar field background value. Figures a) and b) represent allowable regions at different scales. The region with oblique dotted lines  corresponds to allowable values in the case of the system PSR J0737-3039, the horizontal solid lines are the critical value of scalar mass $m_\varphi=2\omega/c$, the vertical dotted lines describe allowable values in the case of the system PSR J1738+0333, the horizontal dot-dash lines correspond to allowable values from the $\gamma_{PPN}$ \citep{lobo}
			.}
		\label{fig:hybridf3}
	\end{figure*}

	It is clear from comparison of constraints (\ref{hybrid2}) and (\ref{hybrid3}) the double binary pulsar gives the better bounds on $\phi_0$ in the hybrid f(R)-gravity. On the other hand, the mixed binary system PSR J1738+0333 provides the best constraints on the scalar field mass. On the Fig.~\ref{fig:hybridf3} we compare the our limits on hybrid f(R)-gravity from binary pulsars with restrictions from the PPN parameter $\gamma_{PPN} = 1 + (2.1 \times 10^{-5}) \pm (2.3 \times 10^{-5})$ \citep{turyshev, lobo}. Thus,  the $\gamma_{PPN}$ gives the better bounds on the $\phi_0$ than the system PSR J0737-3039. The combined restrictions can be obtained from $\gamma_{PPN}$ and system PSR J1738+0333:
	
	\begin{equation}\label{hybrid4} 
	\phi_0\leq0.00004,\ \ \ m_\varphi\leq1.4\times10^{-14} (\text{cm}^{-1}).
	\end{equation} 
	
	The Brans-Dicke model is one of the first scalar-tensor theories which is widespread \citep{bd}. In the framework of this model, many interesting results were obtained including  bounces, wormholes, and constraints on cosmological parameters \citep{rannu, Tretyakova:2011ch, novikov}. After the discovery of the Universe accelerated expansion \citep{acceleration, acceleration1, acceleration2, acceleration3} scientists consider its  massive version as one of the ways to explain this phenomenon \citep{poisson}. The massive Brans-Dicke theory was considered by \citet{massivebd} in the binary pulsars PSR J1012+5307, PSR J1141-6545, and PSR J0737-3039. Authors have shown that the best restrictions on the Brans-Dicke theory are obtained from the Solar System's PPN parameter $\gamma_{PPN}$. This model is thoroughly studied in the binary pulsars and the expression for the orbital period change is already obtained. It is interesting to compare the restrictions on the massive Brans-Dicke theory obtained as a particular case of our more general consideration with the results of \citet{massivebd}. We also add two contributions to the expression of the orbital period decay (PN corrections to the scalar dipole term and scalar dipole-octupole term) unlike the work of \citet{massivebd}. The other difference of our work from paper of \citet{massivebd} is the accounting of the fact that $\partial_0\varphi\neq\partial_r\varphi$ (see equations~(\ref{derivatives phi1}) and (\ref{derivatives phi})). Our approach leads to significant deviations in the final constraints.
	
	The standard massive scalar-tensor gravity can be reduced to the massive Brans-Dicke theory by the following choice of parameters $G_i$ \citep{particular}:
	\begin{equation}\label{bd1} 
	G_2=\cfrac{2\omega_{BD}}{\phi}X+V(\phi),\ G_3=0,\ G_4=\phi,\ G_5=0,
	\end{equation} 
	and hence
	\begin{equation}\label{bd2} 
	G_{4(0,0)}=\phi_0, \ G_{4(1,0)}=1, \ G_{3(1,0)}=0,\ G_{2(0,1)}=\cfrac{2\omega_{BD}}{\phi_0}.
	\end{equation} 
	Here $\omega_{BD}$ is Brans-Dicke parameter.
	
	The dependence of the quantity $\phi_0$ upon $\omega_{BD}$ for massive and massless Brans-Dicke cases \citep{massivebd} is
	\begin{equation}\label{phi0w} 
	\phi_0=\cfrac{4+2\omega_{BD}}{G(3+2\omega_{BD})}.
	\end{equation} 
	
One of the most important feature of the Brans-Dicke model is the presence of sensitivities in structure of the theory \citep{eardley}. Dependence of sensitivity value on a neutron star mass and equations of state was studied in detail earlier by \citet{will, sens}. For our calculations we use approximate values $s_{WD}\sim10^{-4}$ and $s_{NS}\sim 0.2$ \citep{massivebd}. 

Firstly, we obtain the restrictions on the massive Brans-Dicke theory from the double binary pulsar PSR J0737-3039. In the neutron star-neutron star binary $s_a-s_b\approx0$ in massive Brans-Dicke theory \citep{will1, will11, massivebd, will12} and hence scalar dipole, PN corrections to the scalar dipole and scalar dipole-octupole terms vanish. There is only quadrupole terms in expression for restrictions:
	\begin{eqnarray}\label{bd3} 
	1\leq\biggl(\cfrac{3+2\omega_{BD}}{4+2\omega_{BD}}\biggr)^{\frac{5}{3}}&+&\cfrac{0.4(3+2\omega_{BD})^{\frac{2}{3}}}{(2\omega_{BD}+4)^{\frac{5}{3}}}\\
	&-&2\times10^{26}\cfrac{m_\varphi^2(3+2\omega_{BD})^{\frac{2}{3}}}{(2\omega_{BD}+4)^{\frac{5}{3}}}\leq1.031.\nonumber
	\end{eqnarray} 

\begin{figure}
		\begin{minipage}[h]{.45\textwidth}
			\begin{center}
				\includegraphics[width=8cm]{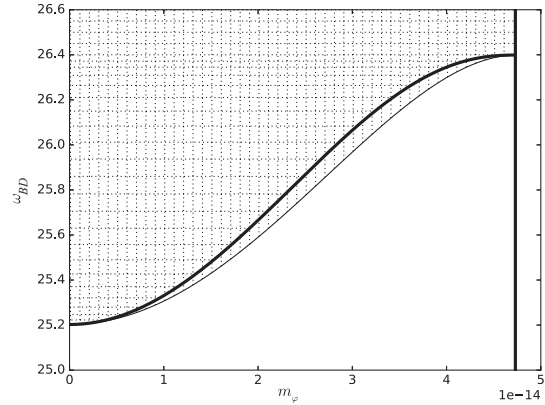}
			\end{center}
			\caption{Massive Brans-Dicke. Dependence of the $\omega_{BD}$ upon the scalar field mass in the case of the system PSR J0737-3039. The vertical dotted lines correspond to the results of \citet{massivebd}. The bold line and region of horizontal dotted lines  describe the results of our work. The vertical solid line is the critical value of scalar mass ($m_\varphi=2\omega/c$).}
			\label{fig:bd0737}
		\end{minipage}
	\end{figure}

	 In the case of neutron star-neutron star system the only deviation of our results from the results of \citet{massivebd} is the account of $\partial_0\varphi\neq\partial_r\varphi$. We reflect on the Fig.~\ref{fig:bd0737} the difference between our and \citet{massivebd} results in the case of PSR J0737-3039. One can see that the difference between two approaches is unimportant in the case of this system.

\begin{figure}
		\begin{minipage}[h]{.45\textwidth}
			\begin{center}
				\includegraphics[width=8cm]{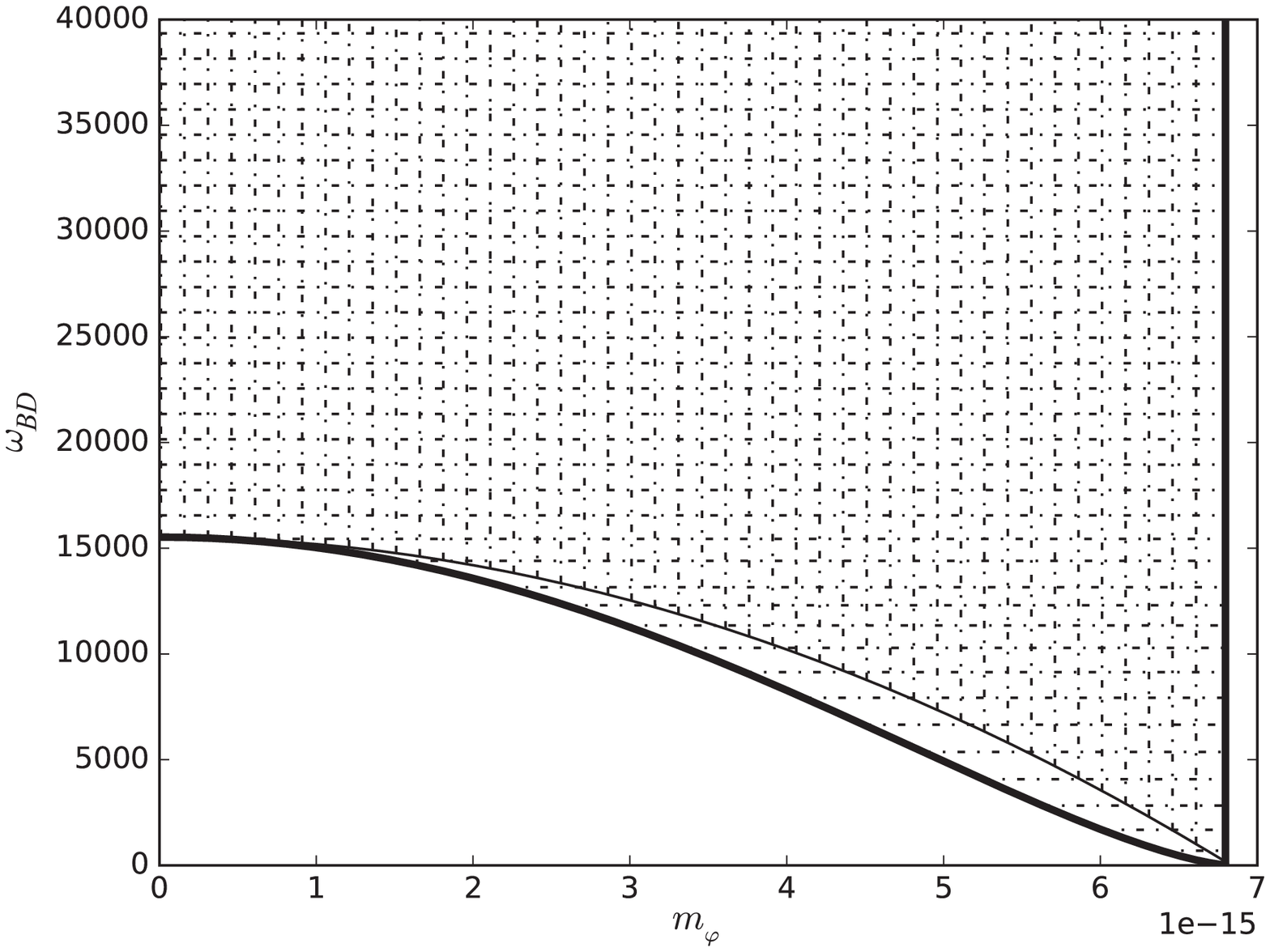}
			\end{center}
			\caption{Massive Brans-Dicke. Dependence of the $\omega_{BD}$ upon the scalar field mass in the case of the system PSR J1738+0333. The solid line andregion of vertical dot-dash lines correspond to the results of \citet{massivebd}. The bold line and region of horizontal dot-dash lines describe the results of our work. The vertical bold line is the critical value of scalar mass ($m_\varphi=\omega/c$).}
			\label{fig:bd1738}
		\end{minipage}
	\end{figure}

	Further we constrain the massive Brans-Dicke gravity using the observational data of the mixed binary system PSR J1738+0333:
	\begin{eqnarray}\label{bd4} 
	1\leq\cfrac{(3+2\omega_{BD})^{\frac{5}{3}}}{(4+2\omega_{BD})^{\frac{5}{3}}}&+&\cfrac{0.4(3+2\omega_{BD})^{\frac{2}{3}}}{(4+2\omega_{BD})^{\frac{5}{3}}}\\
	&+&\cfrac{6\times10^3}{2\omega_{BD}+4}-\cfrac{2\times10^{32}m_\varphi^2}{2\omega_{BD}+4}\leq1.19.\nonumber
	\end{eqnarray} 
	The system was not considered by \citet{massivebd} because the accurate data for this binary pulsar \citep{1738, 17381} has appeared after the publication of \citet{massivebd}. This system predicts the greatest deviations between our approach and the one of \citet{massivebd}. This difference also appears due to the account of $\partial_0\varphi\neq\partial_r\varphi$. The PN corrections to the scalar dipole and scalar dipole-octupole terms introduce insignificant deviations since they are smaller than the dipole term for an order of magnitude. All deviations are reflected on the Fig.~\ref{fig:bd1738}.

	\begin{table}
		\begin{minipage}[!t]{.45\textwidth}
			\caption{Parameters PSR J1012+5307 \citep{kinematic, cal}
			}\label{tab:J1012+5307}
			\begin{tabular}{l l l} 
				\hline
				
				Parameter & Physical meaning & Experimental value\\
				\hline
				$P_b $ & orbital period & $ 0.60467271355(3)$ day\\

				$e$ & eccentricity & $ 0.12(3) \times 10^{-5} $\\

				$\dot P_b^{obs}$ & observational secular & $ 0.50(14) \times 10^{-13}$\\
				& change of $P_b$ &\\
				
				$\dot P_b^{intr}$ & intrinsic secular & $ 0.15(15) \times 10^{-13}$\\
				& change of $P_b$&\\
				
				$\dot P_b^{intr}/\dot P_b^{GR}$ & relation between $\dot P_b^{intr}$& $ 1.36(1.39) $\\
				& and $\dot P_b^{GR}$ &\\
				
				$m_1 $ & mass of the pulsar & $1.64(22)\  M_{\bigodot}$\\
				
				$m_2 $ & mass of the white dwarf & $0.16(2)\  M_{\bigodot}$\\
				
				$m $ & total system mass & $1.8(3)\  M_{\bigodot}
				$\\
				
				\hline
			\end{tabular}
		\end{minipage}
		\end {table}

	Among mixed binary systems considered by \citet{massivebd} the system PSR J1012+5307 is the only one with negligibly small eccentricity \citep{cal, kinematic}. The observational data for this system are reflected in the Table~\ref{tab:J1012+5307}. The mass of white dwarf was obtained from spectroscopic observations and then, using pulsar timing data, mass ratio was found \citep{cal}. From our approach in the mixed binary  PSR J1012+5307 we obtain the following restrictions:
		\begin{eqnarray}\label{bd4} 
		1\leq\cfrac{(3+2\omega_{BD})^{\frac{5}{3}}}{(4+2\omega_{BD})^{\frac{5}{3}}}&+&\cfrac{0.4(3+2\omega_{BD})^{\frac{2}{3}}}{(4+2\omega_{BD})^{\frac{5}{3}}}\\
		&+&\cfrac{8\times10^3}{2\omega_{BD}+4}-\cfrac{7\times10^{32}m_\varphi^2}{2\omega_{BD}+4}\leq4.14.\nonumber
		\end{eqnarray}
		
		The system PSR J1012+5307 gives the best restrictions for the scalar field mass among considering system:
		\begin{equation}\label{massnswd} 
		m_{\varphi}<4\times 10^{-15} (\text{cm}^{-1}).
		\end{equation}
		However, the increasing of the accuracy in determining the scalar field mass, causes the decreasing of accuracy the orbital period change value (\ref{bd4}).
		
\begin{figure}				
				\includegraphics[width=\columnwidth]{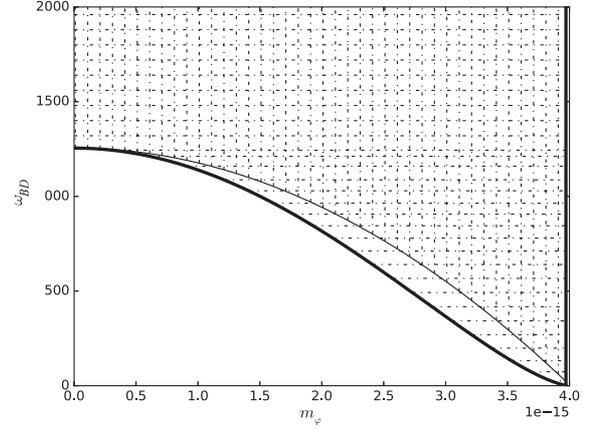}
				\caption{Massive Brans-Dicke. Dependence of the $\omega_{BD}$ upon the scalar field mass in the case of the system PSR J1012+5307. The solid line and region of vertical dot-dash lines correspond to the results of \citet{massivebd}
					. The bold line and region of horizontal dot-dash lines describe the results of our work. The vertical line is the critical value of scalar mass ($m_\varphi=\omega/c$).}
				\label{fig:bd1012}
		\end{figure}

		We choose the system PSR J1012+5307 following to the work of \citet{massivebd}. The deviations between our results and  results of \citet{massivebd} in this system are reflected on the Fig.~\ref{fig:bd1012}. The difference between two approaches is smaller than in  PSR J1738+0333, and restrictions on  $\omega_{BD}$ are worse but ones on the scalar field mass are better. For all parameters both of systems give the better constraints than PSR J0737-3039. In the rest, the conclusions being correct for  PSR J1738+0333 are also true for PSR J1012+5307.
		
		\begin{figure}
				\includegraphics[width=\columnwidth]{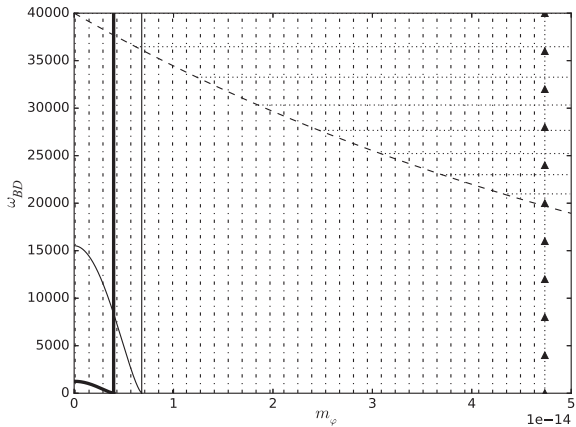}
				\caption{Massive Brans-Dicke. Dependence of the $\omega_{BD}$ upon the scalar field mass. The solid line corresponds to the PSR J1738+0333. The bold line describes the results for the PSR J1012+5307. The line with triangles represents the PSR J0737-3039. The region with vertical dot-dash lines contains the allowable values obtained from binary pulsars. The region with horizontal dotted lines contains the allowable values obtained from Cassini data for $\gamma_{PPN}$ \citep{turyshev}
					.}
				\label{fig:gamma}
		\end{figure}
		
		On the Fig.~\ref{fig:gamma} we present all constraints from binary pulsars with restrictions on the massive Brans-Dicke theory obtained from the PPN parameter $\gamma_{PPN}$ \citep{turyshev}. It is clear that $\gamma_{PPN}$ gives the better constraints on the $\omega_{BD}$ than all of considering binary pulsars. The mixed binary PSR J1012+5307 provides the best bounds on the scalar field mass. Combining the restrictions of these two systems are following:
		\begin{equation}\label{bdall} 
		\omega_{BD}\geq36000,\ \ \ m_\varphi\leq4\times10^{-15} (\text{cm}^{-1}).
		\end{equation}

		\section{Conclusions}
		\label{sec:conclusions}
		
		We considered the subclass of Horndeski gravity without Vainshtein mechanism in the strong field regime of binary pulsars. Our purpose was to impose restrictions on this subclass in the strong field limit. That is why we found the expression for the orbital period decay $\dot P_b^{th}$ predicted by standard massive scalar-tensor theories in quasi-circular orbit approximation. The expression was obtained using the post-Newtonian expansion \citep{will2} of the tensor and scalar fields. The stress-energy pseudotensor was derived using the Noether theorem \citep{saffer}. We showed that the orbital period change consists of two sectors: tensor one and scalar one. 
		
		The tensor sector coincides with the value of the orbital period decay predicted by GR \citep{will1} up to the effective gravitational constant $\mathcal{G}_{ab}$ between two bodies $a$ and $b$. There is neither monopole nor dipole radiations. 
		
		The scalar sector includes monopole, dipole, quadrupole, and dipole-octupole terms. In addition, we considered the PN corrections to the scalar dipole term. The monopole term vanishes in the quasi-circular approximation but all other terms survive. The dipole-octupole contribution adds the negative modification to the scalar energy flux at the same PN order as the scalar quadrupole radiation. 
		
		To impose restrictions on the standard massive scalar-tensor theories in binary pulsars we choose the observational data of mixed binary pulsar PSR J1738-0333 \citep{1738}. This system has the most accurate data among mixed binary pulsars with small eccentricity at the moment. In this system, the scalar dipole term has the leading order in scalar sector. So we focused our attention mostly on this contribution in addition to tensor quadrupole term. 
		
		All terms of scalar sector depend on the propagation speed of the massive scalar mode $v_{\varphi}(\omega, m_\varphi)$. We showed that constraints on the scalar field mass $m_\varphi$ are completely independent of the choice of a specific scalar-tensor theory with the second-order field equations. The scalar field mass depends only on the orbital period of binary pulsar $P_b$. The greater orbital period causes the better scalar field mass restrictions. However, the greater the orbital period provides the worse accuracy of the orbital period decay. The best constraints on the scalar field mass we obtained from the system PSR J1012+5307 but this system has much worse measurement accuracy of the orbital period change \citep{kinematic} than PSR J1738+0333 \citep{1738}. However, the latter loses to the former in the magnitude of the orbital period.
		
		Not all scalar-tensor theories  predict existence of sensitivities.  Regardless of the type of binary pulsar, only quadru\-pole contributions (tensor and scalar) remain in such theories. We showed that in gravitational models without sensitivities the double binary pulsar PSR J0737-3039 gives the best constraints on  all parameters of the theory except the scalar field mass.
		
		We considered contributions of the PN corrections to the scalar dipole term and the scalar cross dipole-octupole term (see also paper of \citet{zhang}). These two terms have the same PN order as scalar quadrupole one therefore they cannot be neglected a priori. However, we showed directly that in two types of binary pulsars contribution of the scalar dipole term and the scalar cross dipole-octupole term are insignificant. 
		
		The important aspect of our work is evidence of injustice of expression $\partial_r\varphi=-\partial_0\varphi$ (see equations~(\ref{derivatives phi1})  and (\ref{derivatives phi})). Earlier the proof was given in the paper for the screened modified gravity \citep{zhang} but we proved injustice this equality using the example of the more general case of scalar-tensor theory.
		
		Also  two specific cases of massive scalar-tensor gravity were considered : the hybrid metric-Palatini f(R)-gravity theory and the massive Brans-Dicke theory. 
		
		The hybrid f(R)-gravity is pure geometrical \citep{hybrid} and there are no sensitivities. Among two pulsars systems PSR J0737-3039 and PSR J1738+0333 the former gives the most accurate restrictions for the background value of scalar field $\phi_0$. Nevertheless, the latter provides the best constraints on the scalar field mass (due to value of the orbital period). We also tested the hybrid f(R)-gravity in the Solar System using the observational data for PPN parameter $\gamma_{PPN}$. From our consideration it is clear that restrictions of $\phi_0$ obtained from $\gamma_{PPN}$ are ahead of accuracy of the constraints obtained from PSR J0737-3039.
		
		The another specific theory, which was considered in our work, is the massive Brans-Dicke model. Earlier this theory studied in detail in binary pulsars by \citet{massivebd} and we were interested in comparison of their and our results. We tested the massive Brans-Dicke theory in three binary pulsars: PSR J0737-3039, PSR J1738+0333, and PSR J1012+5307. The last system was taken into account for direct comparison with the results of \citet{massivebd} and ours. The direct comparison was also carried out for the system PSR J0737-3039. Besides we apply the predication for the orbital period change of  \citet{massivebd} to the system PSR J1738+0333 and compare with our results. We clearly show that the differences are significant especially in the system PSR J1738+0333. The difference between two approaches is that we took into account additional contributions to the scalar sector (the PN corrections to the scalar dipole and the scalar dipole-octupole terms), as well as the injustice of the expression $\partial_r\varphi=-\partial_0\varphi$ (see equations~(\ref{derivatives phi1}  and (\ref{derivatives phi})). The contribution of the PN corrections to the scalar field and the cross term are insignificant, but the equality and inequality of $\partial_r\varphi$ and $-\partial_0\varphi$ leads to essentially different results. Thus we showed by the example that it is necessary take into account the inequality $\partial_r\varphi\neq-\partial_0\varphi$. In addition, we consider the massive Brans-Dicke theory in the Solar System. After comparison of all  constraints our conclusion is the following: the best constraints on the massive Brans-Dicke theory are obtained from mixed test of $\gamma_{PPN}$ (restrictions on $\omega_{BD}$) and PSR J1012+5307 (restrictions on $m_\varphi$).

		As the next step, we plan to investigate the general case of Horndeski theory. One of the most important directions of the further researches is obtaining the expression for sensitivity and determination of it's dependence of star mass in the Horndeski gravity. Also other possible generalization of our work is the test of the Horndeski gravity using all the set of post-Keplerian parameters \citep{damour}.
		
\section*{Acknowledgements}
			
		The authors thank A.N. Petrov for discussions and comments on the topics of this paper. This work was supported by the grant 16-02-00682 from Russian Foundation for Basic Research.

\onecolumn
	\appendix\label{Appendix }
\section{Orbital period change in the massive scalar-tensor theories}
	
	After all the substitutions of parameters $\mathcal{G}_{12}$ from equation~(\ref{geff_1}), and $A_d, \bar{A}_d, A_q, A_o$ from expressions~(\ref{Ad}) and (\ref{adbar}) the first derivative of the orbital period in the standard massive scalar-tensor theories takes the form:
	
	\begin{eqnarray}\label{first derivative of the orbital period} 
	\dot P_b^{th}=&-&\cfrac{192\pi\mu}{5m}\biggl(\cfrac{2\pi m}{c^3G_{4(0,0)}P_b}\biggr)^{\frac{5}{3}}\biggl\{1+\biggl[2c_\varphi G_{4(1,0)}\biggl(1-\cfrac{2s_2}{\phi_0}\cfrac{G_{4(0,0)}}{G_{4(1,0)}}\biggr)-\cfrac{4s_1 c_\varphi G_{4(0,0)}}{\phi_0}\biggl(1-\cfrac{2s_2}{\phi_0}\cfrac{G_{4(0,0)}}{G_{4(1,0)}}\biggr)\biggr]\biggr\}^{\frac{2}{3}}\nonumber\\
	&\times&\Biggl\{1+\cfrac{5c^2c_\varphi G_{4(0,0)}(s_2-s_1)}{24\phi_0}\biggl\{\biggl(\cfrac{2G_{4(0,0)}(s_2-s_1)}{G_{4(1,0)}\phi_0}\biggr)\biggl(\cfrac{P_bG_{4(0,0)}}{2\pi m}\biggr)^{2/3}\biggl(1+\biggl[2c_\varphi G_{4(1,0)}\biggl(1-\cfrac{2s_2}{\phi_0}\cfrac{G_{4(0,0)}}{G_{4(1,0)}}\biggr)\nonumber\\
&-&\cfrac{4s_1 c_\varphi G_{4(0,0)}}{\phi_0}\biggl(1-\cfrac{2s_2}{\phi_0}\cfrac{G_{4(0,0)}}{G_{4(1,0)}}\biggr)\biggr]\biggr)^{-2/3}+\cfrac{2\mu G_{4(0,0)}}{c^2m}\biggl[-\cfrac{7}{2G_{4(0,0)}}\biggl(\cfrac{m_2}{m_1}-\cfrac{m_1}{m_2}\biggr)+\cfrac{1}{G_{4(1,0)}\phi_0}\biggl(\cfrac{7m_2s_1}{m_1}-\cfrac{7m_1s_2}{m_2}\nonumber\\
&+&6s_1-6s_2\biggr)+\cfrac{23}{4}c_\varphi\cfrac{G_{4(1,0)}}{G_{4(0,0)}}\biggl(\cfrac{m_2}{m_1}-\cfrac{m_1}{m_2}\biggr)+\cfrac{c_\varphi}{\phi_0}\biggl(\cfrac{7m_1s_2}{m_2}-\cfrac{7m_2s_1}{m_1}+\cfrac{23m_1s_1}{2m_2}-\cfrac{23m_2s_2}{2m_1}+4s_1-4s_2\biggr)\nonumber\\
&+&\cfrac{G_{4(0,0)}c_\varphi}{G_{4(1,0)}\phi_0^2}\biggl(\cfrac{14s_1s_2m_2}{m_1}-\cfrac{14s_1s_2m_1}{m_2}+8s_1-8s_2+\cfrac{8m_2s_1}{m_1}-\cfrac{8m_1s_2}{m_2}+\cfrac{9s^2_2m_1}{m_2}-\cfrac{9s^2_1m_2}{m_1}-\cfrac{8m_2s_1'}{m_1}+\cfrac{8m_1s_2'}{m_2}-8s_1'\nonumber\\
&+&8s_2'+8s_2^2-8s_1^2\biggr)+\cfrac{G^2_{4(0,0)}c_\varphi}{G^2_{4(1,0)}\phi_0^3}\biggl(\cfrac{18s^2_1s_2m_2}{m_1}-\cfrac{18s^2_2s_1m_1}{m_2}+16s^2_1s_2-16s^2_2s_1+\cfrac{16m_1}{m_2}-\cfrac{16m_2}{m_1}-\cfrac{16s_1s_2'm_1}{m_2}\nonumber\\
&+&\cfrac{16s_2s_1'm_2}{m_1}-16s_1s_2'+16s_2s_1'\biggl)\biggr]\biggl(1+\biggl[2c_\varphi G_{4(1,0)}\biggl(1-\cfrac{2s_2}{\phi_0}\cfrac{G_{4(0,0)}}{G_{4(1,0)}}\biggr)-\cfrac{4s_1 c_\varphi G_{4(0,0)}}{\phi_0}\biggl(1-\cfrac{2s_2}{\phi_0}\cfrac{G_{4(0,0)}}{G_{4(1,0)}}\biggr)\biggr]\biggr)^{-1}\biggr\}\nonumber\\
&\times& \biggr[1-\biggl(\cfrac{ P_bc m_\varphi}{2\pi}\biggr)^2\biggr]^{\frac{3}{2}}+ \cfrac{G_{4(1,0)}c_\varphi}{3}\biggl(1-\cfrac{2G_{4(0,0)}(s_2m_1+s_1m_2)}{G_{4(1,0)}m\phi_0}\biggr)^2\biggr[1-\biggl(\cfrac{ P_bc m_\varphi}{4\pi}\biggr)^2\biggr]^{\frac{5}{2}}-\cfrac{c_\varphi}{96}\biggl(\cfrac{2G_{4(0,0)}(s_2-s_1)}{\phi_0}\biggr)\nonumber\\
&\times&\biggl(\cfrac{m_1-m_2}{m}-\cfrac{2G_{4(0,0)}}{G_{4(1,0)}\phi_0}\cfrac{s_2m_1^2-s_1m_2^2}{m^2}\biggr)\biggr[1-\biggl(\cfrac{ P_bc m_\varphi}{2\pi}\biggr)^2\biggr]^{\frac{5}{2}}\Biggr\}
	\end{eqnarray}

\bsp	
\label{lastpage}

\begin{thebibliography}{99}
			\bibitem[\protect\citeauthoryear{Abbott et al.}{2017}]{grb}
		Abbott B. P. et al. (LIGO Scientific Collaboration and Virgo Collaboration),  2017, Astrophys. J. Lett.,  848, L13
			\bibitem[\protect\citeauthoryear{Abbott et al.}{2017}]{abbott}
		Abbott B. P.  et al. (LIGO Scientific Collaboration and Virgo Collaboration),  2017, Phys. Rev. Lett., 119, 161101 
			\bibitem[\protect\citeauthoryear{Alexeyev \& Pomazanov}{1997}]{Alexeev:1996vs}
		Alexeyev S., Pomazanov M., 1997,  Phys. Rev. D,  55, 2110 
			\bibitem[\protect\citeauthoryear{Alexeyev, Rannu \& Gareeva}{2011}]{rannu}
		Alexeyev S. O.,  Rannu K. A., Gareeva D. V., 2011, J. Exp. Theor. Phys, 113, 4, 628 
			\bibitem[\protect\citeauthoryear{Alexeyev \& Rannu}{2012}]{Alexeyev:2012zz}
		Alexeyev S. O., Rannu K. A.,  2012, J. Exp. Theor. Phys., 114, 406
			\bibitem[\protect\citeauthoryear{Alsing et al.}{2012}]{massivebd}
		Alsing J., Berti E., Will C. M.,  Zaglauer H.,  2012, Phys. Rev. D,  85, 064041 
			\bibitem[\protect\citeauthoryear{Antoniadis et al.}{2012}]{1738}
		Antoniadis J. et al.,  2012, Mon. Not. R. Astron. Soc.,  423, 4, 3316
			\bibitem[\protect\citeauthoryear{Archibald et al.}{2018}]{archi}
		Archibald A. M. et al., 2018, Nature, 559, 73
			\bibitem[\protect\citeauthoryear{Arnoulx de Pirey Saint Alby \& Yunes}{2017}]{Arnoulx}
		Arnoulx de Pirey Saint Alby T., Yunes N., 2017,  Phys. Rev. D 96, 064040
			\bibitem[\protect\citeauthoryear{Ashtekar, Bonga \& Kesavan}{2016}]{ashtekar}
		Ashtekar A.,  Bonga B., Kesavan A.,  2016, Phys. Rev. Lett., 116, 051101
			\bibitem[\protect\citeauthoryear{Babichev \& Esposito-Far\'ese}{2013}]{babichev}
		Babichev E., Esposito-Far\'ese G., 2013, Phys. Rev. D 87, 044032
			\bibitem[\protect\citeauthoryear{Baker et al.}{2017}]{Baker}
		Baker T. et al., 2017, Phys. Rev. Lett. 119, 251301
			\bibitem[\protect\citeauthoryear{Bertotti, Iess \& Tortora}{2003}]{turyshev}
		Bertotti B., Iess L., Tortora P.,  2003, Nature, 425, 374 
			\bibitem[\protect\citeauthoryear{Bhat, Bailes \& Verbiest}{2008}]{bhat}
		Bhat N. D. R.,  Bailes M.,  Verbiest J. P. W., 2008, Phys. Rev. D, 77, 124017
			\bibitem[\protect\citeauthoryear{Boisseau}{2011}]{poisson}
		Boisseau B., 2011,  Phys.Rev. D, 83, 043521 
			\bibitem[\protect\citeauthoryear{Borka Jovanovic et al.}{2016}]{dark matter1}
		Borka Jovanovic V., Capozziello S.,  Jovanovic P.,  Borka D., 2016, Phys. Dark  Univ., 14, 73
			\bibitem[\protect\citeauthoryear{Brans \& Dicke}{1961}]{bd}
		Brans C., Dicke H.,  1961, Phys. Rev.,  124, 925 
			\bibitem[\protect\citeauthoryear{Brax, Davis \& Sakstein}{2014}]{brax}
		Brax P.,  Davis A.-C.,  Sakstein J., 2014, Class. Quantum Grav., 31, 225001
			\bibitem[\protect\citeauthoryear{Burgay et al.}{2003}]{kramer}
		Burgay M. et al., 2003,  Nature, 426, 531 
			\bibitem[\protect\citeauthoryear{Callanan, Garnavich \& Koester}{1998}]{cal}
		Callanan P. J.,  Garnavich P. M., Koester D., 1998, Mon. Not. R. Astron. Soc., 298, 207 
			\bibitem[\protect\citeauthoryear{Capozziello et al.}{2013}]{hybrid}
		Capozziello S. et al.,  2013, J. Cosmol. Astropart. Phys., 1304, 011 
			\bibitem[\protect\citeauthoryear{Capozziello et al.}{2013}]{dark matter3}
		Capozziello S., Harko T., Koivisto T. S.,  Lobo F. S. N., Olmo G. J., 2013,  J. Cosmol. Astropart. Phys., 1307, 024 
			\bibitem[\protect\citeauthoryear{Capozziello et al.}{2015}]{hybrid1}
 		Capozziello S. et al.,  2015, Univ., 1, 2, 199 
			\bibitem[\protect\citeauthoryear{Clifton et al.}{2012}]{Clifton}
		Clifton T., Ferreira P. G. , Padilla A., Skordis C., 2012, Physics Reports, 513, 1, 1
			\bibitem[\protect\citeauthoryear{Damour \& Deruelle}{1985}]{damour}
		Damour T.,  Deruelle N., 1985, Ann. Inst. Henri Poincare A, 43, 107 
			\bibitem[\protect\citeauthoryear{Damour \& Deruelle}{1986}]{damour1}
 		Damour T.,  Deruelle N., 1986, Ann. Inst. Henri Poincare A, 44, 263 
			\bibitem[\protect\citeauthoryear{Damour \& Esposito-Far\'ese}{1992}]{damour3}
  		Damour T., Esposito-Far\'ese G., 1992,  Class. Quant. Grav., 9, 2093
			\bibitem[\protect\citeauthoryear{Damour \& Esposito-Far\'ese}{1996}]{damour3}
  		Damour T., Esposito-Far\'ese G., 1996,  Phys. Rev. D, 53, 5541 
			\bibitem[\protect\citeauthoryear{Damour \& Polyakov}{1994a}]{polyakov}
		Damour T.,  Polyakov A. M., 1994, Nucl. Phys. B 423, 532 
			\bibitem[\protect\citeauthoryear{Damour \& Polyakov}{1994b}]{polyakov2}
		 Damour T., Polyakov A. M., 1994, General Relativity and Gravitation 26, 1171 
			\bibitem[\protect\citeauthoryear{Damour \& Taylor}{1991}]{damourtaylor}
		Damour T., Taylor J. H.,  1991, Astrophys. J., 366, 501
			\bibitem[\protect\citeauthoryear{Damour \& Taylor}{1992}]{damour2}
		Damour T., Taylor J. H.,  1992, Phys.Rev. D, 45, 1840 
			\bibitem[\protect\citeauthoryear{De Felice \& Tsujikawa}{2012}]{dark energy in horndeski}
		De Felice A., Tsujikawa S., 2012,  J. Cosmol. Astropart. Phys., 1202, 007 
			\bibitem[\protect\citeauthoryear{Desvignes  et al.}{2016}]{pulsars}
		Desvignes G. et al.,  2016, Mon. Not. R. Astron. Soc., 458, 3, 3341 
			\bibitem[\protect\citeauthoryear{Di Casola, Liberati \& Sonego}{2015}]{casola}
		Di Casola E., Liberati S., Sonego S., 2015, Am. J. Phys., 83, 39 
			\bibitem[\protect\citeauthoryear{Eardley}{1975}]{eardley}
		Eardley D. M., 1975, Astrophys. J. Lett., 196, L59 
			\bibitem[\protect\citeauthoryear{Eddington}{1922}]{will2}
		Eddington A. S., 1922, The Mathematical Theory of Relativity. Cambridge University Press, London
			\bibitem[\protect\citeauthoryear{Einstein,  Infeld \&  Hoffmann}{1938}]{eih}
		Einstein A.,  Infeld L., Hoffmann B.,  1938, Ann. Math., 20, 39, 65 
			\bibitem[\protect\citeauthoryear{Esposito-Far\'ese}{2011}]{Esp}
 		Esposito-Far\'ese G., 2011, Fundam.Theor.Phys. 162, 461
			\bibitem[\protect\citeauthoryear{Ezquiaga \& Zumalacarregui}{2017}]{ez}
		Ezquiaga J. M., Zumalacarregui M.,  2017, Phys. Rev. Lett., 119, 251304 
			\bibitem[\protect\citeauthoryear{Freire, Kramer \& Wex}{2012a}]{2012a}
		Freire P. C. C., Kramer M., Wex N., 2012a, Classical and Quantum Gravity, 29, 184007
			\bibitem[\protect\citeauthoryear{Freire et al.}{2012b}]{17381}
 		Freire P. C. C. et al., 2012b, Mon. Not. R. Astron. Soc.,  423, 4, 3328
			\bibitem[\protect\citeauthoryear{Fujii \& Maeda}{2003}]{Fujii}
		Fujii Y.,  Maeda K., 2003, The scalar-tensor theory of gravitation. Cambridge University Press, Cambridge, England
			\bibitem[\protect\citeauthoryear{Galiautdinov \& Kopeikin}{2016}]{kopeikin}
		Galiautdinov A.,  Kopeikin S. M., 2016, Phys. Rev. D 94, 044015 
			\bibitem[\protect\citeauthoryear{Gao}{2011}]{gao}
		Gao X., 2011, J. Cosmol. Astropart. Phys., 1110, 021 
			\bibitem[\protect\citeauthoryear{Germani \&  Martin-Moruno}{2017}]{germani}
		Germani C., Martin-Moruno P.,  2017, Phys. Dark Univ., 18, 1 
			\bibitem[\protect\citeauthoryear{Hinterbichler \& Khoury}{2010}]{hint}
		Hinterbichler K., Khoury J., 2010, Phys. Rev. Lett. 104, 231301 
			\bibitem[\protect\citeauthoryear{Hinterbichler et al.}{2011}]{hint1}
		Hinterbichler K., Khoury J.,  Levy A., Matas A., 2011, Phys. Rev. D84, 103521 
			\bibitem[\protect\citeauthoryear{Hohmann}{2015}]{hohmann}
		Hohmann M., 2015, Phys. Rev. D, 92, 064019 
			\bibitem[\protect\citeauthoryear{Horndeski}{1974}]{horndeski}
		Horndeski G. W., 1974,  Int. J. Theor. Phys., 10, 363 
			\bibitem[\protect\citeauthoryear{Hou \& Gong}{2018}]{hou}
		Hou S., Gong Y.,  2018, Eur. Phys. J. C, 78, 247 
			\bibitem[\protect\citeauthoryear{Hulse \& Taylor}{1975}]{hulse}
		Hulse R.  A.,  Taylor J. H., 1975, Astrophys. J. Lett., 195, L51 
			\bibitem[\protect\citeauthoryear{Ivanov, Pshirkov \& Rubtsov}{2016}]{Ivanov:2016ifg}
		Ivanov M., Pshirkov M.,  Rubtsov G.,  2016, Phys. Rev. D, 94, 063004 
			\bibitem[\protect\citeauthoryear{Katsuragawa \& Matsuzaki}{2017}]{dark matter2}
 		Katsuragawa T., Matsuzaki S.,  2017, Phys. Rev. D, 95, 044040
			\bibitem[\protect\citeauthoryear{Kennedy, Lombriser \& Taylor}{2017}]{germani1}
		Kennedy J., Lombriser L., Taylor A.,  2017, Phys.Rev. D, 96, 084051
			\bibitem[\protect\citeauthoryear{Khoury \& Weltman}{2004a}]{khw}
 		Khoury J., Weltman A., 2004, Phys. Rev. D 69, 044026
			\bibitem[\protect\citeauthoryear{Khoury \& Weltman}{2004b}]{khw1}
 		Khoury J., Weltman A., 2004, Phys. Rev. Lett. 93, 171104
			\bibitem[\protect\citeauthoryear{Kimura,  Kobayashi \&  Yamamoto}{2012}]{kimura}
		Kimura R., Kobayashi T., Yamamoto K., 2012,  Phys. Rev. D,  85, 024023 
			\bibitem[\protect\citeauthoryear{Kobayashi,  Yamaguchi \& Yokoyama}{2011}]{kobayashi}
		Kobayashi T., Yamaguchi M., Yokoyama J.,  2011, Prog. Theor. Phys., 126, 511 
			\bibitem[\protect\citeauthoryear{Koyama,  Niz \& Tasinato}{2013}]{koyama}
		Koyama K., Niz G., Tasinato G., 2013, Phys. Rev. D, 88, 021502
			\bibitem[\protect\citeauthoryear{Kramer et al.}{2006}]{kramer1}
 		Kramer M. et al.,  2006, Science, 341, 97 
			\bibitem[\protect\citeauthoryear{Lazaridis et al.}{2009}]{kinematic}
		Lazaridis K. et al., 2009,  Mon. Not. R.Astron.Soc.,  400, 2, 805 
			\bibitem[\protect\citeauthoryear{Leanizbarrutia,  Lobo \&  Saez-Gomez}{2017}]{lobo}
		Leanizbarrutia I.,  Lobo F. S. N., Saez-Gomez D., 2017, Phys. Rev. D, 95, 084046 
			\bibitem[\protect\citeauthoryear{McManus, Lombriser \&  Penarrubia}{2016}]{particular}
		McManus R.,  Lombriser L., Penarrubia J., 2016, J. Cosmol. Astropart. Phys., 11, 006, 1611 
			\bibitem[\protect\citeauthoryear{Morse \& Feshbach}{1953}]{morse}
		Morse P. M., Feshbach H., 1953,  Methods of Theoretical Physics. McGraw-Hill, New York
			\bibitem[\protect\citeauthoryear{Narikawa et al.}{2013}]{narikawa} 
		Narikawa T., Kobayashi T., Yamauchi D., Saito R., 2013,  Phys.Rev. D, 87, 124006 
			\bibitem[\protect\citeauthoryear{Nordtvedt}{1968}]{will21}
		Nordtvedt K.,  1968, Phys. Rev., 169, 1017 
			\bibitem[\protect\citeauthoryear{Novikov et al.}{2014}]{novikov}
		Novikov I. D.,  Shatskiy A. A., Alexeyev S. O., Tretyakova D. A., 2014,  Phys. Usp., 57, 352
			\bibitem[\protect\citeauthoryear{Nunes,  Martn-Moruno \& Lobo}{2017}]{germani2}
 		Nunes N.J., Martn-Moruno P., Lobo F.S.N., 2017,  Univ., 3, 33 
			\bibitem[\protect\citeauthoryear{Oort}{1932}]{Zwicky1}
  		Oort J. H.,  1932, Bull. Astron. Inst. Neth., 6, 249
			\bibitem[\protect\citeauthoryear{Perlmutter et al.}{1999}]{acceleration2}
		Perlmutter S. et al.,  1999, Astrophys. J., 517, 565
			\bibitem[\protect\citeauthoryear{Petrov}{2008}]{petrov}
		Petrov A. N., 2008, Classical and Quantum Gravity Research. Nova Science Publishers, New York
			\bibitem[\protect\citeauthoryear{Poisson \& Will}{2014}]{will12}
 		Poisson  E., Will C. M., 2014, Gravity: Newtonian, Post-Newtonian, Relativistic. Cambridge University Press, London
			\bibitem[\protect\citeauthoryear{Pshirkov, Tuntsov \& Postnov}{2008}]{pshirkov}
		Pshirkov M., Tuntsov A., Postnov K., 2008, Phys. Rev. Lett., 101, 261101 
			\bibitem[\protect\citeauthoryear{Ransom et al.}{2014}]{ransom}
		Ransom S. M. et al., 2014, Nature, 505, 520
			\bibitem[\protect\citeauthoryear{Renk, Zumalacarregui \& Montanari}{2016}]{salvatelli1}
		Renk J., Zumalacarregui M., Montanari F., 2016, J. Cosmol. Astropart. Phys., 1607, 040 
			\bibitem[\protect\citeauthoryear{Riess et al.}{1999}]{acceleration} 
		Riess A. G.  et al., 1999,  Astron. J., 116, 1009
			\bibitem[\protect\citeauthoryear{Riess et al.}{2004}]{acceleration1}
		Riess A. G. et al., 2004, Astrophys. J., 607, 665
			\bibitem[\protect\citeauthoryear{Saffer, Yunes \& Yagi}{2018}]{saffer}
		Saffer A.,  Yunes N., Yagi K., 2018, Class. Quantum Grav., 35, 5, 055011
			\bibitem[\protect\citeauthoryear{Sakstein}{2014}]{Sakstein}
		Sakstein J., 2014, J. Cosmol. Astropart. Phys., 1412, 012
			\bibitem[\protect\citeauthoryear{Salvatelli, Piazza \& Marinoni}{2016}]{salvatelli}
		Salvatelli V., Piazza F., Marinoni C., 2016, J. Cosmol. Astropart. Phys., 1609, 027 
			\bibitem[\protect\citeauthoryear{Shi, Li \& Han}{2017}]{dark matter}
		Shi D.,  Li B., Han J., 2017,  Mon. Not. R. Astron. Soc., 469, 705
			\bibitem[\protect\citeauthoryear{Spergel et al.}{2007}]{acceleration3}
 		Spergel D. N. et al., 2007, Astrophys. J. Suppl. Ser., 170, 377
			\bibitem[\protect\citeauthoryear{Stairs}{2005}]{stairs}
		Stairs I. H. et al., 2005, Astrophys. J., 632, 1060
			\bibitem[\protect\citeauthoryear{Taylor \& Weisberg}{1982}]{tw}
		Taylor J. H., Weisberg J. M. 1982, Astrophys. J., 253, 908
			\bibitem[\protect\citeauthoryear{Teyssandier \& Tourranc}{1983}]{tey}
		Teyssandier P., Tourranc P.,  1983, J. Math. Phys.,  24, 2793 
			\bibitem[\protect\citeauthoryear{Tretyakova et al.}{2012}]{Tretyakova:2011ch} 
		Tretyakova D. A.,  Shatskiy A. A.,  Novikov I. D., Alexeyev S. O., 2012, Phys. Rev. D, 85, 124059
			\bibitem[\protect\citeauthoryear{Tretyakova}{2017}]{Dasha1}
		Tretyakova D. A.,  2017, J. Exp. Theor. Phys., 125, 403
			\bibitem[\protect\citeauthoryear{Tretyakova \& Latosh}{2018}]{Dasha}
		Tretyakova D. A.,  Latosh B. N.,  2018, Univ., 4, 2, 26
			\bibitem[\protect\citeauthoryear{Turner}{1999}]{dark energy}
		Turner M. S., 1999,  The Third Stromlo Symposium: The Galactic Halo, 165, 431
			\bibitem[\protect\citeauthoryear{Vainshtein}{1972}]{Vainshtein}
		Vainshtein A. I., 1972, Phys. Lett. B, 39, 393 
			\bibitem[\protect\citeauthoryear{Will}{1971}]{will22}
  		Will C. M.,  1971, Astrophys. J., 163, 611 
			\bibitem[\protect\citeauthoryear{Will \&  Nordtvedt}{1972}]{will23}
		 Will C. M., Nordtvedt K.,  1972, Astrophys. J., 177, 757 
			\bibitem[\protect\citeauthoryear{Will}{1981}]{will1}
		Will C. M.,  1981,  Theory and Experiment in Gravitational Physics. Cambridge University Press, London
			\bibitem[\protect\citeauthoryear{Will \& Zaglauer}{1989}]{will}
		Will C. M., Zaglauer H. W.,  1989, Astrophys.J., 346, 366 
			\bibitem[\protect\citeauthoryear{Will}{2014}]{will11}
		Will C. M., 2014, Liv. Rev. Relat.,  17, 4 
			\bibitem[\protect\citeauthoryear{Zaglauer}{1992}]{sens}
		Zaglauer H.,  1992, Astrophys.J.,  393, 685 
			\bibitem[\protect\citeauthoryear{Zhang, Liu \& Zhao}{2017}]{zhang}
		Zhang X., Liu T., Zhao W., 2017, Phys. Rev. D, 95, 104027 
			\bibitem[\protect\citeauthoryear{Zwicky}{1933}]{Zwicky}
		Zwicky F., 1933, Helv. Phys. Acta, 6, 110 
	\end{thebibliography}
\end{document}